\documentclass[11pt, a4paper]{article}
\usepackage[lmargin=1.5cm,rmargin=1.5cm,tmargin=1.5cm,bmargin=1.5cm]{geometry} 
\usepackage{amsbsy}
\usepackage{amsmath}
\usepackage{graphicx}
\usepackage{mdframed}
\usepackage[format=hang, margin=15pt]{subfig}
\usepackage{microtype} 
\usepackage[labelfont=bf]{caption}
\usepackage{booktabs} 
\usepackage{paralist} 
\usepackage{authblk}
\usepackage{float}
\usepackage{url}

\usepackage{listings}
\usepackage{natbib}
\usepackage[utf8]{inputenc}
\usepackage{xcolor}
\usepackage{hyperref} 
\hypersetup{backref=true,       
    pagebackref=true,               
    hyperindex=true,                
    colorlinks=true,                
    breaklinks=true,                
    urlcolor= black,                
    linkcolor= blue,                
    bookmarks=true,                 
    bookmarksopen=false,
    filecolor=black,
    citecolor=blue,
    linkbordercolor=blue
}
\graphicspath{{figs/}{../figs/}}

\usepackage{fancyhdr}
\fancypagestyle{plain}{
         \fancyhf{}
         \fancyfoot[R]{\footnotesize\thepage}
}

\usepackage{caption}
\usepackage{lipsum}
\makeatletter
\renewcommand{\maketitle}{\bgroup\setlength{\parindent}{0pt}
\begin{flushleft}
  \vspace{0.1cm}

  \textbf{\@title} 
  \vspace{0.3cm}

  \@author 
  \vspace{0.3cm}

 \@date
\end{flushleft}\egroup
}
\makeatother

\title{{\large Guidance note on best statistical practices for TOAR analyses}} 
\author[1]{Kai-Lan Chang} 
\author[2]{Martin G. Schultz}
\author[3]{Gerbrand Koren} 
\author[2]{Niklas Selke}
\affil[1]{Cooperative Institute for Research in Environmental Sciences, University of Colorado Boulder and NOAA Chemical Sciences Laboratory, Boulder, CO, USA}
\affil[2]{J\"ulich Supercomputing Centre, Forschungszentrum J\"ulich, Germany}
\affil[3]{Copernicus Institute of Sustainable Development, Utrecht University, Utrecht, the Netherlands}
\date{This document has been produced based on discussions in the statistics focus working group of the TOAR-II initiative, and approved by the TOAR-II Steering Committee\\  \vspace*{0.1mm} April 26, 2023}

\begin{document}
\begin{mdframed}
  \noindent\raisebox{-\height}{\includegraphics[width=6cm]{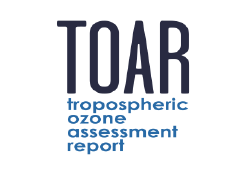}}\hfill
  \begin{minipage}[t]{\dimexpr \textwidth-6cm-\columnsep}
     \maketitle
  \end{minipage}
\end{mdframed}

The aim of this guidance note is to provide recommendations on best statistical practices and to ensure consistent communication of statistical analysis and associated uncertainty across TOAR publications. The scope includes approaches for reporting trends, a discussion of strengths and weaknesses of commonly used techniques, and calibrated language for the communication of uncertainty. The focus of this guidance note is placed on trend analysis, which is expected to be the main statistical topic of interest across many TOAR-II focus working groups, but some of the recommendations and principles provided below are also valid for other applications. Recommendations are highlighted and numbered from R1 to R9.

\section{Definition of trend analysis}
The purpose of trend analysis can be defined as
\begin{verbatim} 
    detecting and attributing the change and its uncertainty of the 
    statistical properties in a time series of a predefined variable.
\end{verbatim}
where the statistical properties represent how we define and express the trend (e.g. mean, median or extreme percentile changes). Each component of this definition will be discussed through relevant topics and recommendations as follows: Section 2 clarifies the distinct roles of linear and nonlinear techniques in trend quantification. Section 3 compares the commonly adopted statistical properties in trend assessment. Section 4 discusses data preparation for time series analysis. Section 5 identifies the fundamental factors affecting uncertainty quantification associated with the trend estimate. Section 6 demonstrates how the reliability of trend assessment can be determined through the relationship between quantified trend and uncertainty. Section 7 discusses the change point detection of trends. Section 8 provides additional resources.

This guidance note aims to provide recommendations of tropospheric ozone trend analysis from a practical aspect, without discussing technical details, although we do provide some R and Python code for quantile regression in the Annex E. Practitioners (referring to authors of publications written in the context of the TOAR activity) who are interested in the fundamental theory of trend analysis (e.g. the statistical relationship between the sample size, magnitude of trend and uncertainty) are referred to various textbooks \citep{chandler2011, von2001, wilks2011} and articles \citep{chang2021, weatherhead1998}.

\section{Quantitative versus qualitative summary of the trend}
\begin{mdframed}
\begin{flushleft}
\begin{itemize}
\item[R1:] Practitioners are recommended to provide quantitative trends and should avoid drawing conclusions when the magnitude of trends and associated uncertainties are not explicitly quantified.
\end{itemize}
\end{flushleft}
\end{mdframed}

Trend detection techniques can be generally classified as linear or nonlinear methods. By using a large number of parameters and numerical optimization, nonlinear methods offer a great deal of flexibility in modeling the nonlinear pattern of a time series. However, the result is often too complex mathematically to quantitatively provide a simple trend value and an uncertainty value. Unless there is a clear understanding of (non-linear) processes driving the trend, quantitative non-linear trend analysis poses the risk of overfitting. Under this circumstance, nonlinear trends can be visualized and summarized qualitatively. We provide a further critical discussion of nonlinear trend methods with an example in Annex A for interested readers. Apparent changes in the magnitude of the trend, which are often the motivation to apply non-linear methods, can usually be handled well through piecewise linear trend analysis in combination with change point detection (see Section 7). 

Due to mathematical challenges, we have limited options to quantitatively summarize the trends. The common statistics for reporting linear trends include the mean (e.g. from the least squares method), median (e.g. from the least absolute deviations method or Sen-Theil method), and a range of percentiles (e.g. from the quantile regression method). These statistics represent different data characteristics of trends and will be discussed more thoroughly in the next section. 

To alleviate the disconnect between the issues of linearity and general nonlinearity, previous studies have chosen to conduct trend analysis by using both nonlinear and linear trend techniques \citep[e.g.,][]{cooper2020, oltmans2006}; the former is used to visually inspect the general nonlinearity, and investigate the appropriateness of the linear assumption, and the latter aims to provide a quantitative statement and acknowledge the uncertainty. Therefore, the application of linear trends should not imply our belief in a strictly linear assumption, rather, we have chosen this approach in order to communicate with and convey messages to a more general audience \citep{crimmins2020}. 

It is possible to model nonlinearity in a linear additive regression model. For short-term nonlinearity, we recommend incorporating meteorological or relevant covariates into regression-based methods to attribute the data variability (see Section 4 in \cite{chang2021} for a demonstration). For long-term nonlinear change, employing piecewise linear trends for change point analysis is preferable \citep{banerjee2020, weatherhead1998}, since the trends and uncertainties before and after the change point can be explicitly quantified (see Section 7 and Annex A). The advantage of this attribution and detection approach is that the result is fully and quantitatively interpretable. Nevertheless, this exercise requires an iterative fitting process, because the best set of covariates and the location of the change point(s) need to be identified by the practitioners (see Annex B for further discussions), and the best fitted model often varies for different time series.

\begin{mdframed}
 \begin{flushleft}
\begin{itemize}
\item[R2:] Visualize data series and/or try non-linear trend detection methods to assess the validity of a linear trend model before reporting linear trends quantitatively.
\item[R3:] When time series exhibit clearly visible changes in the magnitude of a trend over time, we recommend the application of piecewise linear trends in combination with a change point detection algorithm.
\end{itemize}
\end{flushleft}
\end{mdframed}

\section{Recommendation for trend analysis technique and comparison of statistical properties across common trend techniques}

\begin{mdframed}
 \begin{flushleft}
\begin{itemize}
\item[R4:] In TOAR-II, we recommend quantile regression (QR) for trend analysis.
\end{itemize}
\end{flushleft}
\end{mdframed}

Briefly, the reasons are: 1) QR is a well-suited technique for detecting heterogeneous distributional changes \citep{koenker2001}, which is often the case for free tropospheric and surface ozone as they typically show diverse percentile trends; and 2) QR is a regression-based method that allows us to incorporate covariates for attributing data variability and piecewise trends for change point analysis. The main limitation to QR is that a larger sample size is required for a valid estimation of extreme variations, which means that datasets with small or insufficient sample sizes should be limited to analysis of the median trend.

More broadly, this section will also provide the basic context regarding the commonly applied linear trend techniques, so researchers can be reminded of the implicit assumptions behind the methods and what challenges may be encountered during the manuscript peer-review process. Here we select three broadly used techniques, including generalized least squares (GLS), the Sen-Theil method (which was adopted in TOAR-I, but abandoned for TOAR-II; as a candidate of median trend estimates, it can be replaced by QR which has additional advantages discussed below), and QR (which has been incorporated into the TOAR-II database, and is recommended for use). Table 1 gives a brief comparison of statistical capabilities from different techniques, which can be addressed in two aspects: 

\begin{itemize}
\item \textit{The distinctions between means, medians and other percentiles}: The relevant principles are listed in the purple section of Table 1. We can see these principles are closely related to the concept of sample medians (e.g. robust to outliers, small sample size, and non-normal error distribution) or sample means (e.g. having a unique solution, and making use of all data points). It is worth emphasizing that there is a contradiction between making use of all information and resisting the influence of outliers. For example, the Sen-Theil estimator was shown to tolerate arbitrary corruption of up to 29\% of the data points without deterioration of its estimation \citep{wilcox2011}. However, this property also means that up to 29\% of all data are automatically ignored and treated the same as ``non well-behaved" data, even if they are valid samples. Being robust to outliers is not always superior when we have proper data quality control (see Annex C for further discussions), because there is no reason to waste the information about the extreme values when those data are deemed to be valid. Nevertheless, under this circumstance quantile regression can still account for all extreme events by providing a range of percentile trend estimates, instead of merely focusing on the mean or median trends.

\item \textit{The distinctions between regression-based methods and classical non-parametric methods}: The advantages of regression-based approaches over the classical methods (e.g. Sen-Theil) are listed in the orange section of Table 1. All of these features are considered to be necessary or desirable for trend analysis, i.e., accounting for autocorrelation, incorporating covariates to attribute data variability, or integrating adaptive components to study nonlinearity.
\end{itemize}

\begin{table}
      \begin{minipage}{\textwidth}
\caption{A brief comparison of statistical capabilities of three commonly applied techniques.}
\begin{center}
\includegraphics[width=0.7\textwidth]{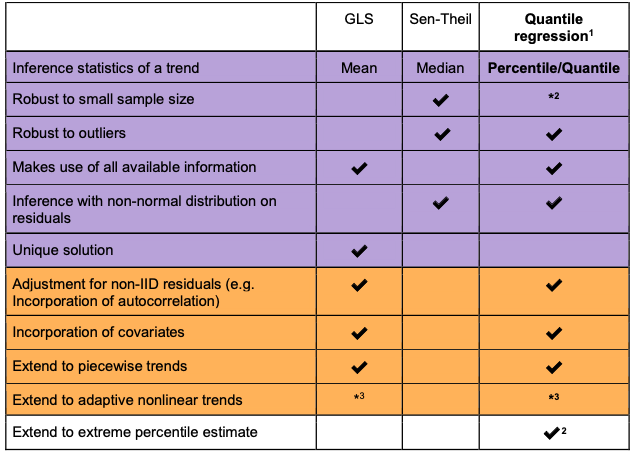}
\end{center}
 \vspace*{-7mm} 
{\footnotesize 
\begin{itemize}
  \setlength\itemsep{0em}
\item[\textsuperscript{1}] Least absolute deviations method (median regression) can be considered to be a special case of quantile regression, so it is not listed independently.
\item[\textsuperscript{2}] For quantile regression, the median estimate is robust to the small sample size problem, but a larger sample size is required for extreme percentile estimates \citep[depending on applications,][]{das2019}. See Table 2 for TOAR-specific recommendations.
\item[\textsuperscript{3}] These extensions can be made via the class of standard or quantile generalized additive models.
\end{itemize}}
      \end{minipage}
\end{table}

Table 1 provides a limited comparison and ignores certain unorthodox extensions. For example, 1) one can combine bootstrapping techniques with the GLS method so the results are more robust to outliers and small sample size problems, and 2) the idea of Sen-Theil estimator can be used to solve multiple regression models using an iterative process \citep{wilcox2011}, but these are beyond our current scope. By focusing on the big picture, the items in Table 1 represent the basic methodology, and can be used as a reference for the strengths and weaknesses of each method. Based on this comparison the researchers can also better understand the hidden context behind the trend detection techniques. Indeed, for the practitioners who applied GLS for their trend analysis, the least squares method is frequently questioned for its lack of robustness to outliers. However, the Sen-Theil method is rarely criticized for its under-appreciation of data records, since it ignores a large portion of extreme values without identifying true outliers from other valid data (a "true" outlier is a high leverage/influence data point that should be identified by specialized knowledge, instead of solely by analytical tools). Familiarity with those principles will help the practitioner make stronger arguments and avoid inappropriate or erroneous statistical applications.

To facilitate comparisons of trends and uncertainties across the TOAR analyses, it is desirable to define the default quantiles to be estimated. Table 2 provides rule-of-thumb recommendations on default percentile trends to be reported according to various ranges of sample size (i.e. the number of data points fitted in the trend model, not raw data). To avoid over-interpretation, we do not recommend reporting percentile trends in intervals narrower than 5\%.

\begin{mdframed}
\begin{flushleft}
\begin{itemize}
\item[R5:] Choose the specified quantiles in QR regression depending on the number of samples.
\end{itemize}
\end{flushleft}
\end{mdframed}

\begin{table}
\caption{Recommendation on default quantiles to be reported with respect to different sample sizes.}
\begin{center}
\includegraphics[width=0.7\textwidth]{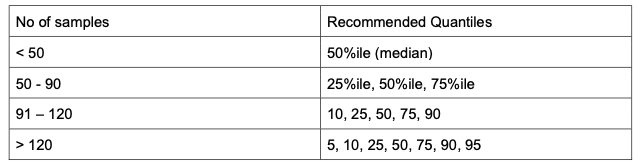}
\end{center}
 \vspace*{-7mm} 
{\footnotesize Analyses of extreme quantiles (minima, 1\%, 2\%, 98\%, 99\%, maxima) require special care and detailed evaluation of the statistical model and diagnostic. In fact, the generalized extreme value (GEV) or threshold exceedance models are designed for the purpose of studying the very extreme percentiles and should be preferred \citep{gilleland2020b, krock2022}.}
\end{table}

\section{Data preparation for trend analysis}

\begin{mdframed}
\begin{flushleft}
\begin{itemize}
\item[R6:] Seasonal adjustments are required for time series analysis. 
\end{itemize}
\end{flushleft}
\end{mdframed}

Before a demonstration of QR is made, it is desirable to briefly discuss data preparation for trend analysis. Conventional regression analysis (the least squares method) requires that the error term has a constant variance. Therefore analyses of atmospheric composition data have been primarily carried out using monthly, seasonal, or annual aggregated values. The rationale behind such aggregations is simply to reduce heterogeneous data variability and meet the homoscedasticity assumption (but monthly values are considered to be less heterogeneous than daily values, only if the size of aggregated samples is large enough or the sample frequency is appropriate). Nevertheless, from a modern statistical perspective, some important information might be lost after data aggregation, which limits our understanding regarding the causes of the data variability (see Annex Figure A6 and \cite{chang2023} for further discussions). We thus recommend preserving as much information as possible by utilizing QR to detect heterogeneous changes.

Ozone and many other atmospheric variables have distinct seasonal and/or diurnal cycles. These known variations should be included in the statistical model that is used to derive trend estimates (see Annex C), or the data should be deseasonalized prior to the trend analysis. Failure to do so will lead to unreasonably large uncertainty estimates and may even adversely impact on the trend estimator (for example, if a time series starts at a seasonal minimum and ends at a seasonal maximum). The requirement for seasonal adjustments was originally proposed in the field of economics \citep[see a review by][]{hannan1970}, since then seasonal adjustments have become a necessary step in time series analysis. 

\textit{Demonstration: a comparison of trend analyses for ozone data at Mace Head, Ireland:}\\
The code and technical details for deseasonalization are provided in Annexes E and F, respectively. Here the time series is deseasonalized in advance. Now we demonstrate the applicability of QR (For comparisons with other methods, we use the monthly time series here, but QR can be properly applicable to daily values too, see Annex F): Figure 1 compares the percentile trend estimates derived by QR to a range of broadly used mean/median trend methods \citep{chang2021, yu2017}, based on the monthly surface ozone anomalies measured at Mace Head, Ireland \citep{cooper2020}. We can clearly observe the unique feature of QR for deriving trends in addition to the central tendency. It should be noted that even though the different methods may show similar trend values, the uncertainty estimators can vary greatly (as shown in the comparison of uncertainties from GLS, Sen-Theil and QR). The main reason for an unrealistically narrow uncertainty from the Sen-Theil estimator is its assumption that residuals are independent and identically distributed (IID, see Section 5 for further discussions). Applying other reasonable techniques for estimating the mean or median change is not forbidden or even discouraged, but we recommend QR be used as the standard for trend analyses in TOAR, because it is a well-designed technique for studying heterogeneous ozone change and quantifying uncertainty across a wide range of percentiles.

\begin{figure*}
\begin{center}
\includegraphics[width=0.7\textwidth]{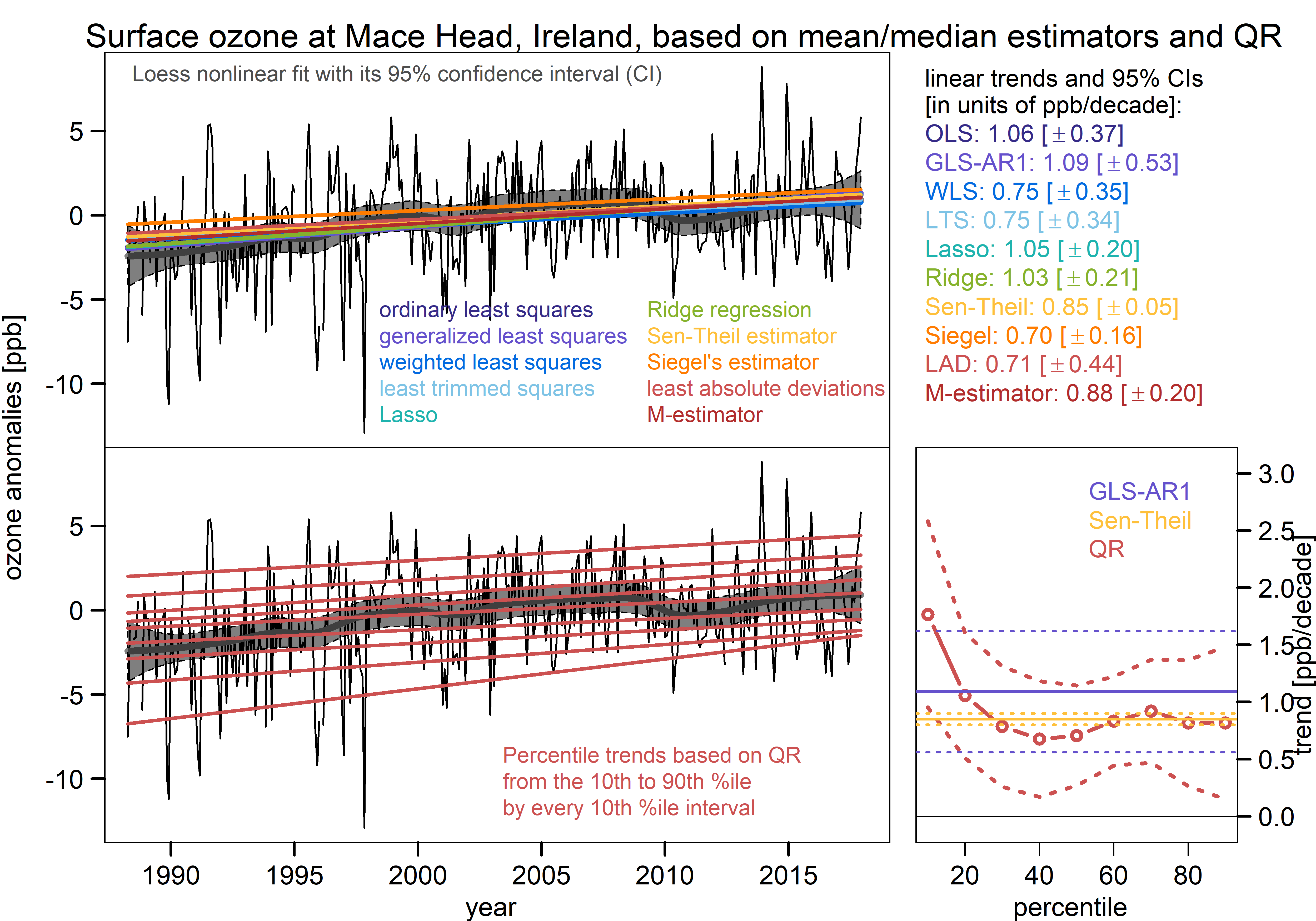}
\end{center}
 \vspace*{-5mm} 
\caption{A demonstration of the difference between a range of mean/median trend methods (upper panel) and percentile trends derived from QR (lower panel), based on surface ozone anomalies measured at Mace Head, Ireland. \\
\footnotesize{Loess (locally estimated scatterplot smoothing, gray curve) fit is added to show nonlinearity of the sample means.  Dashed lines in the lower right panel represent the 95\% confidence intervals, which are computed through the Wilcox test for the Sen-Theil estimator \citep{wilcox2001} and a Huber sandwich estimate for QR \citep{koenker1999}. Note that LAD (least absolute deviations) is a special case of QR when only the median trend is considered.}}
\end{figure*}

\section{Derivation of trend uncertainty}

\begin{mdframed}
\begin{flushleft}
\begin{itemize}
\item[R7:] Practitioners should provide associated uncertainty for any statistical estimation, typically the 95\% confidence interval.
\end{itemize}
\end{flushleft}
\end{mdframed}
 
One of the most critical components of statistical analysis is to acknowledge the uncertainty. Every estimation must be accompanied by a quantification of the associated uncertainty (or error bar), which is used to assess the reliability of the (trend) estimate and is considered to be as equally important as the estimate \citep{chang2021, cressie2021, tong2019}. In terms of linear trend analysis, the uncertainty is referred to as the standard error (SE) of the trend value. The higher the SE, the larger the uncertainty for the trend estimate. Most programming languages provide a summary of fitted results (see Figure 2 for examples of output from Python and R), which at a minimum consists of estimate, uncertainty, $t$-value and $p$-value for each parameter. The $t$-value is the ratio between the estimate and uncertainty (also known as signal-to-noise ratio or SNR), which is used to determine the $p$-value. The interpretation of the reliability of a trend based on $p$-value or SNR is provided in Section 6. 

\begin{figure*}
\includegraphics[width=0.7\textwidth]{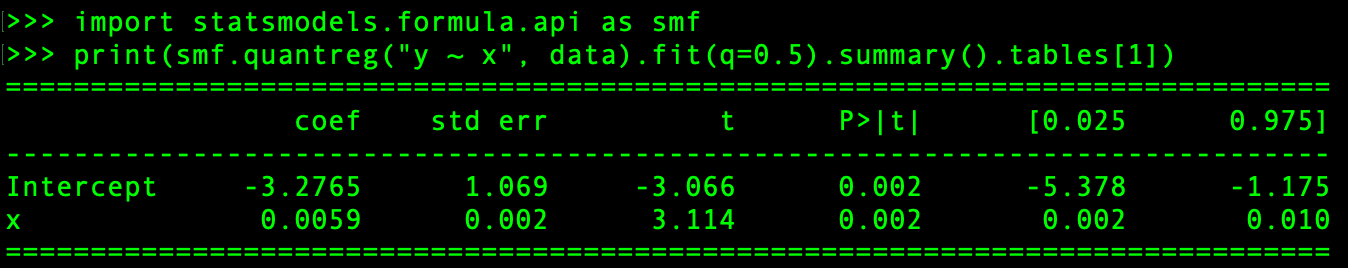}\\
\includegraphics[width=0.4\textwidth]{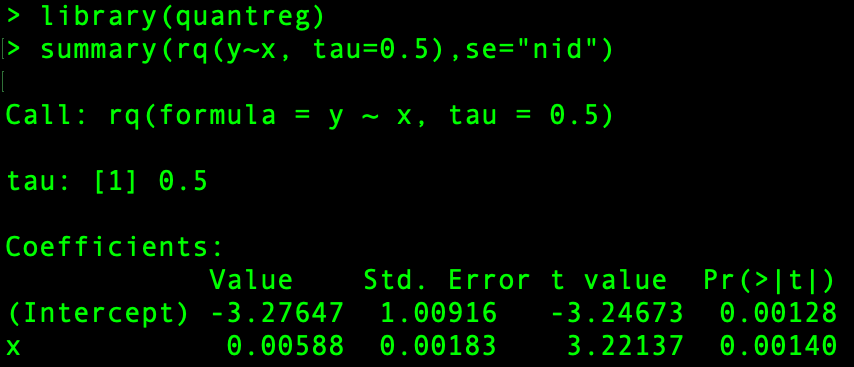}
\caption{Output of summary statistics from quantile/median regression in Python (upper) and R (lower), based on ozone anomalies ($y$) measured at Mace Head, Ireland (as shown in Figure 1). The estimate and standard error (SE) for linear trend ($x$) is in units of ppbv/month. \\
\footnotesize{Note that the default method for quantifying SE in Python is based on the theoretical asymptotic distribution approximated through a kernel estimator \citep{hall1988}. R does not provide a default method for quantifying SE and it has to be specified by the user (here \lstinline{se="nid"} indicates an non-IID SE computed through a Huber sandwich estimate \citep{koenker1999}). Both methods account for heteroskedasticity in the residuals, so the resulting uncertainties are similar. More options for estimating SE are available for Python and R, but these options are designed for either IID or heteroskedasticity cases, and currently none of the options explicitly address autocorrelation. }}
\end{figure*}

There are many factors affecting the uncertainty value. In general, 1) the larger sample size, 2) the smaller data variability, and/or 3) the weaker autocorrelation and heteroskedasticity, result in a smaller uncertainty of a trend. Whereas the first two factors do not directly involve the model assumptions or formulation, IID is generally too strong an assumption because autocorrelation/heteroskedasticity is common in atmospheric composition time series, so the practitioners are recommended to take non-IID residuals into account. The common approaches include incorporation of an autoregressive model \citep{weatherhead1998}, adoption of the heteroskedasticity and autocorrelation consistent SE \citep{kiefer2002}, prewhitening \citep[a process to remove autocorrelation from data,][]{yue2002}, or inference with block bootstrapping methods \citep{politis2004}, otherwise the resulting uncertainty is likely to be inappropriate or underestimated. Due to mathematical challenges, among these approaches, only block bootstrap resampling and prewhitening methods are currently applicable to all trend techniques (Here we only focus on moving or circular block bootstrap. While prewhitening is generally applicable, it was shown to drastically distort data structure in some cases \citep{razavi2018}). The usual resampling methods are unlikely to preserve autocorrelation structure, since they do not sample a series of consecutive data in the time series. Moving block bootstrap fixes this issue by making inference from several blocks of sampled data, instead of from individual random samples \citep{fitzenberger1998, lahiri2007}. Circular (moving) block bootstrap further extends the methodology by wrapping around the start and end points of the time series, so the data near the beginning and end of the record are not systematically under-sampled \citep{gilleland2020a}.

\begin{figure*}
\begin{center}
\includegraphics[width=0.7\textwidth]{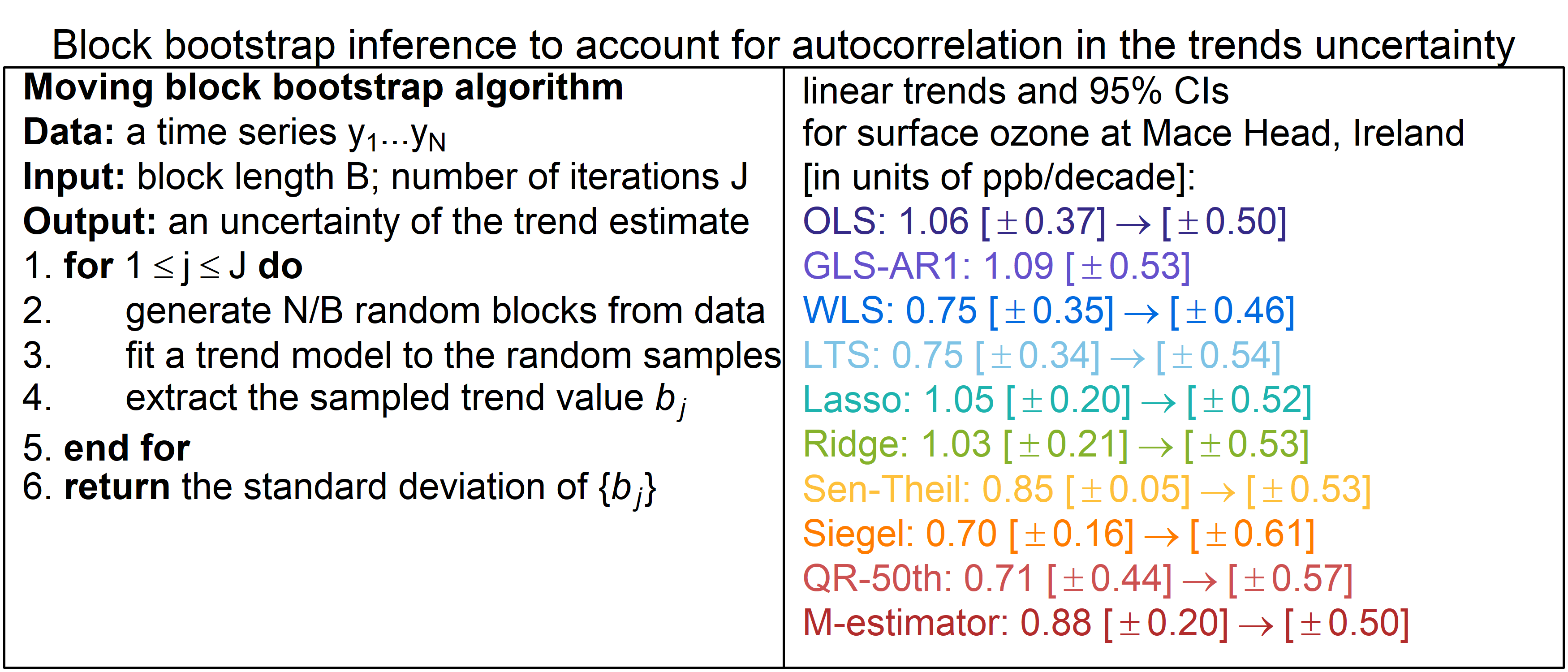}
\end{center}
 \vspace*{-5mm} 
\caption{Code outline for the moving block bootstrap algorithm for deriving the trend uncertainty from autocorrelated time series, and its impact on the trend uncertainty for different trend techniques.\\
\footnotesize{Results are based on a block length $B=N^{0.25}$ (i.e. each selected block contains $B$ consecutive values), number of samples $J$=1000, and ozone data at Mace Head, Ireland. Block bootstrapping was not applied to GLS-AR1 as it already accounts for autocorrelation. Note that QR-50th is equivalent to LAD (least absolute deviations).}}
\end{figure*}

In general, this bootstrapping approach assumes that the uncertainty based on statistical theory can be replaced by empirical uncertainty (meaning that this approach is not limited by some items listed in Table 1, as long as the data are representative). Figure 3 shows the code outline for the moving block bootstrap algorithm, tailored to the purpose of taking autocorrelation into account in the trend uncertainty. The circular block bootstrap algorithm can be adopted by simply wrapping the time series. The optimal block length is often difficult to determine, here we adopt $B=N^{0.25}$ (the power of the sample size to a quarter) as suggested by \cite{hall1995}. We then apply this algorithm to adjust the trend uncertainty for the ozone anomaly time series measured at Mace Head, Ireland (only GLS considers autocorrelation in Figure 1, with the AR1 coefficient of 0.34, and therefore moving block bootstrap is not applied). We can see the uncertainty increased for all techniques after autocorrelation is accounted for, and all methods now have not only similar trend values, but also similar (appropriate) uncertainties. 
 
In terms of nonlinear trend techniques, since they do not provide an overall trend estimate, the 95\% confidence intervals apply to the entire fitted curve, which represents the uncertainty range of the sample means (and should not be interpreted as the 95\% range of the data variability, see Figure 1 for a demonstration).

\section{Calibrated language for communicating uncertainty} 

\begin{mdframed}
\begin{flushleft}
\begin{itemize}
\item[R8:] Gradated calibrated language should be used as a replacement for dichotomous ``statistical significance".
\end{itemize}
\end{flushleft}
\end{mdframed}  

In cases when the trend and its uncertainty need to be reported for a time series, and/or when the trends need to be compared from multiple sites or multiple products, it is desirable to use an uncertainty scale for assessing the reliability and likelihood of the estimated trend, instead of merely dichotomizing the outcome as significant or insignificant \citep{wasserstein2016}. A comprehensive guidance note for communicating uncertainties in terms of confidence, likelihood and probability was previously developed for the IPCC Fifth Assessment Report \citep{mastrandrea2010}; here we adopt a few simplifications (only placing the focus on the level of reliability for trend assessment and with fewer likelihood scales) to allow for consistent and precise communication regarding trend reliability across TOAR analyses (Table 3). Trend reliability can be expressed with the p-value \citep{wasserstein2019} or the signal-to-noise (SNR) ratio \citep{chang2021}.

\begin{table}
\caption{Trend reliability scale}
\begin{center}
\includegraphics[width=0.7\textwidth]{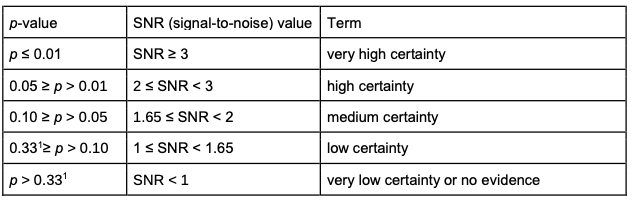}
\end{center}
 \vspace*{-7mm} 
{\footnotesize \textsuperscript{1} This boundary is meant to be fuzzy around 1/3 \citep{mastrandrea2010}.}
\end{table}

This scale is similar to the scheme used in the first phase of TOAR \citep{fleming2018, gaudel2018, mills2018}, but adds another gradation for ``very high certainty", approximately corresponding to a $p$-value $\leq$ 0.01 or SNR $\geq$ 3, which allows us to investigate ``the world beyond $p < 0.05$" \citep{wasserstein2019} and isolate the results with greater statistical confidence for further discussions. For the outcomes with even greater likelihood, the authors can consider directly using the SNR values as a reference. Importantly, this uncertainty scale allows us to avoid using the expression ``statistically significant", which is now widely recognized as being problematic and not useful \citep{wasserstein2019}.

\section{Change point detection}

\begin{mdframed}
\begin{flushleft}
\begin{itemize}
\item[R9:] Apply change point detection to validate any suspicious changes of data structure.
\end{itemize}
\end{flushleft}
\end{mdframed}
 
Change point analysis is highly relevant to the detection of trend change and attribution of intervention (e.g. evaluation of effectiveness of a measure to reduce emissions). A change point does not have a unique meaning in the literature, it can be a break or turning point connecting two data series that are considered to have different averages \citep[a constant shift,][]{weatherhead2017}, variances \citep[distinct variabilities,][]{beaulieu2012} and/or trends \citep{chen2011}. Although these structural changes can be identified through numerical algorithms \citep{killick2014, muggeo2008}, it is up to practitioners to ensure that the change point of trends is not induced by inconsistent or incompatible data \citep[e.g. a mean shift caused by new instruments,][]{weatherhead2017}. A visual inspection of time series anomaly plots can help us to identify different scenarios. 

Factors affecting the detection of trends (e.g. seasonality and autocorrelation) are relevant to the detection of change points, so deseasonalization and incorporation of necessary components in the regression model are suggested. The change point algorithms are generally not robust to small perturbations in the data (meaning that similar change points can be repeatedly detected due to small scale meteorological variations). Given ozone's high temporal and spatial variability, practitioners are recommended to consider the change point analysis of trends only when data are sufficiently long (e.g. no more than one meaningful change point of trends within a decade).

Examples of the change point analysis of a time series with a turnaround of trends is provided in Annex A, and with an abrupt change in Annex B. Based on a scientific point of view, when a change point of data averages or trends is identified and attributable, such as new air pollution control policies taking effect or changes of site environment (e.g. shutdown of a factory), it is not due to natural variability and therefore should be addressed by the trend model (some studies refer to this as the intervention analysis, see \cite{weatherhead1998}). If the cause of a change point cannot be associated with any specific factors, practitioners are recommended to investigate the time series at nearby stations (see Annex C for further discussions) and/or compare to chemistry-climate model simulations.

\section{Additional Resources}
A step-by-step demonstration of trend analysis using quantile regression (with Python and R code) can be found in Annex E. A recording of an introductory lecture describing quantile regression can be found in \cite{chang2022a}.

\section*{Acknowledgments}
KLC was supported in part by NOAA cooperative agreements NA17OAR4320101 and NA22OAR4320151. NS and MGS acknowledge the European Research Council and the European Union to have funded the ERC Advanced Grant 787576 (IntelliAQ). We would like to thank Owen R. Cooper, Erika von Schneidemesser, Eric Gilleland, Daan Hubert, Arno Keppens and Davide Putero for helpful and thought-provoking suggestions which improved the content of this work.

\clearpage
\appendix
\setcounter{section}{0} \renewcommand{\thesection}{Annex \Alph{section}}
\setcounter{table}{0} \renewcommand{\thetable}{A\arabic{table}}
\setcounter{figure}{0} \renewcommand{\thefigure}{A\arabic{figure}}

\section{Discussion on the differences between the ways to capture nonlinearity and the ways to summarize trends} 
There are many statistical methods that can be used to study the nonlinearity of a trend, such as moving averages, polynomials, Loess (locally estimated scatterplot smoothing), smoothing splines, dynamical linear modeling, vector autoregressive modeling, and empirical mode decomposition. With the exception of polynomials, these methods are considered to be the adaptive approach, i.e. no assumptions of the shape/structure of trends have to be made in advance \cite[a series of references on this topic is provided by][]{chang2021}. However, these techniques are often too complex mathematically (e.g. involving too many hyperparameters) to quantitatively provide a simple trend value and an uncertainty value. 

Based on the above discussions, there is a large disconnect between the ways to study nonlinearity and the ways to summarize the trends. We demonstrate this disconnect in Figure A1, using monthly surface ozone anomalies measured from the mountaintop site of Zugspitze, Germany (1978-2020) as an example \citep{cooper2020}. The major feature of this time series is that from a visual inspection a clear overall increase of ozone can be observed between 1978 and the late 1990s, and the increasing trend appears to diminish afterward. In this figure three different methods are used to investigate the trends; even though these methods are able to capture the diminishment of the trends, their interpretations can be quite different:

\begin{itemize}
\item \textit{Adaptive learning methods}: Here the Loess locally weighted regression is selected to identify the multi-year fluctuations in the time series, but it does not provide a quantitative estimate of the trend. Most adaptive methods have the capability to be tuned to capture the local nonlinearity (and their local 95\% uncertainty range can be estimated), so the resulting trends can be highly irregular/fluctuating (as shown by the blue Loess curve). The rationale of this smoothness/roughness tuning is to remove as much short-term (unstructured) variation as possible, while maintaining the long-term tendency (in simple terms, the goal is to find a balance between data fidelity and model complexity). Nevertheless, as noted previously, the nonlinear trends derived from the adaptive methods are often too complex to condense into a simple equation, so the results can only be summarized qualitatively, instead of quantitatively. For example, the blue curve in Figure A1 should be interpreted as follows: \textit{the long-term change can be observed as an increasing trend between 1978 to the late 1990s, and a relatively flat trend from 1998 to 2020, with some short-term variations attached}. 

\item \textit{Piecewise linear segments}: In this approach each line segment represents a linear trend tailored to a specific period (preferably for a sufficiently long term), so any nonlinearity of the resulting trends will only occur when the practitioner deems it necessary. In contrast to adaptive learning methods, the specifications of this approach need to be customized by the practitioner (e.g. identifying any possible change point(s), determining any data discontinuity due to data instability issues, etc). Nevertheless, since the practitioner fully controls the specifications, they are able to explicitly and quantitatively summarize the trend value and its uncertainty for each linear segment. For example, using the Loess smoother as guidance, we select 1998 as the change point of the trends, the red curve in Figure A1 represents \textit{a positive slope of 0.049 [$\pm$0.008] ppbv/month (or 5.91 [$\pm$1.04] ppbv/decade, p $\leq$ 0.01, very high certainty) over 1978-1998 and a negative slope of -0.007 [$\pm$0.007] ppbv/month (or -0.81 [$\pm$0.88] ppbv/decade, p = 0.07, medium certainty) over 1998-2020}. It should also be noted that even though we only specify two linear segments to represent the long-term change, the short-term variations can be accounted for by incorporating relevant covariates, such as meteorological variables into a multiple linear regression \citep{chang2021, wells2021}.

\item \textit{Polynomials}: While the use of polynomials may be appealing due to their simplicity of calculation (any spreadsheet can fit a polynomial to a time series), except for a specific scenario when the practitioner aims to study the acceleration/deceleration rate of the changes \citep{paolo2015}, they are not recommended for the general purpose of ozone trend detection. This approach is an intermediate compromise between adaptive methods and piecewise linear trends. As can be seen in Figure A1 (orange curve), the disadvantages of an in-between method can be obvious: 1) Polynomials are not as flexible as adaptive methods for properly capturing shorter term interannual variations, and higher order polynomials can result in a very poor fit due to a lack of measures to prevent numerical issues at boundary conditions (see the supplement to \cite{cooper2020} for further details and examples); 2) To facilitate interpretation of the results, the applications of polynomial trends are mostly limited to the orders of 2 or 3, out of a desire to keep the equation simple and avoid overfitting. However, the polynomial equations do not directly provide any interpretable trend and associated uncertainty values (in contrast to the piecewise method). While the polynomial fit in Figure A1 minimizes the distance between the data points and the curve, it does not provide a realistic representation of the long-term data variability and trends. The Loess smoother shows a peak of ozone in 1998 and a relative minimum several years later in 2008; in contrast the polynomial implies an increase since 1978 that reaches a peak in 2007. The reason for the incorrect peak location is due to the fact that a simple 2nd order polynomial has to bend symmetrically about its vertex, regardless of the heterogeneous changes of the data variability; this symmetrical bend also gives the false impression that ozone has steadily decreased since 2007, even though the Loess smoother shows a small increase. Another problem common to polynomials is that as the time series extends into the future, the fit to the early part of the time series and the peak of the trend typically shift and vary in intensity, an unrealistic scenario that does not occur when using a stable curve fitting technique such as the Loess smoother (see the supplement to \cite{cooper2020} for further details and examples). 
\end{itemize}

\begin{figure*}
\begin{center}
\includegraphics[width=0.5\textwidth]{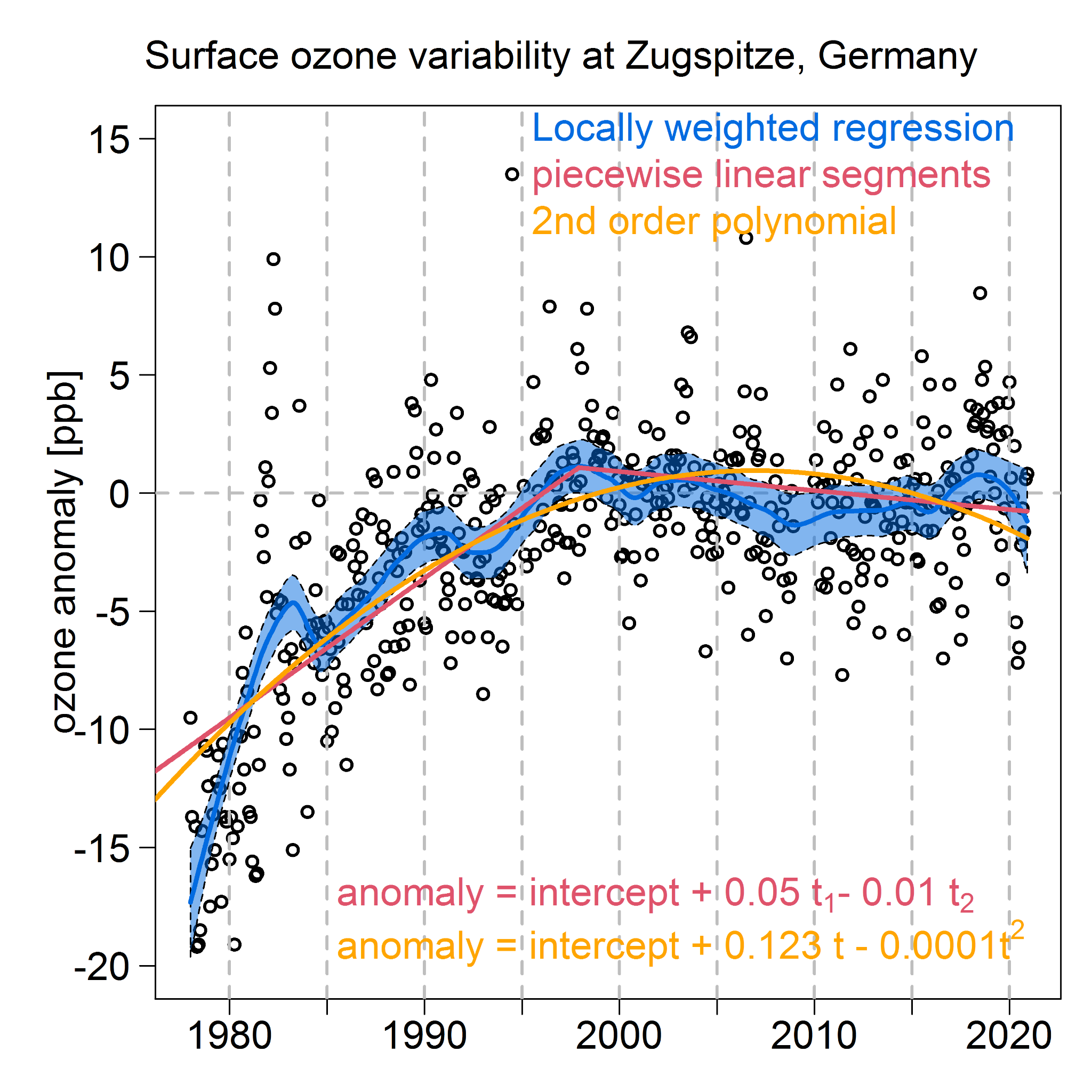}
\end{center}
 \vspace*{-5mm} 
\caption{Comparison of trend fitting techniques from Loess locally weighted regression (with 95\% confidence interval), piecewise linear segments and second order polynomials, based on the monthly nighttime ozone anomalies measured from Zugspitze, Germany (1978-2020), where $t_1$ and $t_2$ in the equation represent the segment before and after the change point (thus the coefficients are the trends in units of ppbv/month), respectively, and $t$ is the collection of $t_1$ and $t_2$ representing the temporal index of the time series.}
\end{figure*}

\section{Advice on change point analysis of trends}
\textit{Searching for the location of change points}: While Figure A1 adopted piecewise linear segments to carry out change point analysis, it may leave the readers unaware of how to select the change point in an objective way (when the change point is difficult to determine visually). The common approach is to find the optimal location of a change point, such that either the fitted residuals are minimized, or the difference in magnitude of trends or SNRs before/after the change point is maximized. It should be noted that the exact change point is not critical, as long as the rough locations can be identified (e.g. the year is known, but the month is unknown, or unimportant). Once the location of the change point is found, we can evaluate if the model with the change point outperforms the model without the change point, based on various model performance metrics \citep{reeves2007}. However, it might be infeasible to implement such a complex algorithm across a large database such as TOAR (e.g. over ten thousand time series). To simplify the methodology, we may conclude that a change point is valid if the difference in magnitude of SNRs before and after the change point is maximized and exceeds a certain threshold (e.g. $\text{dSNR} = |\text{SNR(before)} - \text{SNR(after)}| \geq 2$). 

Although various learning methods were proposed to automatically identify multiple change points \citep{baranowski2019, verbesselt2010}, one should be cautious when a change point is found near the beginning or end of the data record, and when multiple change points are detected within a short period, since the time frame might be too short to conclude that the long-term trend has changed.

\textit{Discussion on the change point of trends and an abrupt change}: For a demonstrative purpose, let us consider a phenomenon known as Simpson's paradox: When we combine data series from two consecutive periods that both show a positive trend, is it possible to produce an overall negative trend? This result is shown in Figure A2, based on surface ozone anomalies measured at Centennial, Wyoming (1990-2021, US EPA Clean Air Status and Trends Network or CASTNET). We can clearly observe an abrupt change in spring of 2004. If a linear trend is fitted to the entire period, a negative slope is found (-0.46 [$\pm$0.69] ppbv/decade, $p=0.18$, low certainty). Nevertheless, as indicated by a mean shift model (i.e. only two constant terms are fitted and no trend is considered), a clear shift of data averages is observed between 1990-2004 and 2004-2021 (-1.97 [$\pm$1.15] ppbv, $p \leq 0.01$). So this change point can be identified as a mean shift, which suggests that the overall negative trend is induced by the abrupt change. 

\begin{figure*}
\begin{center}
\includegraphics[width=0.5\textwidth]{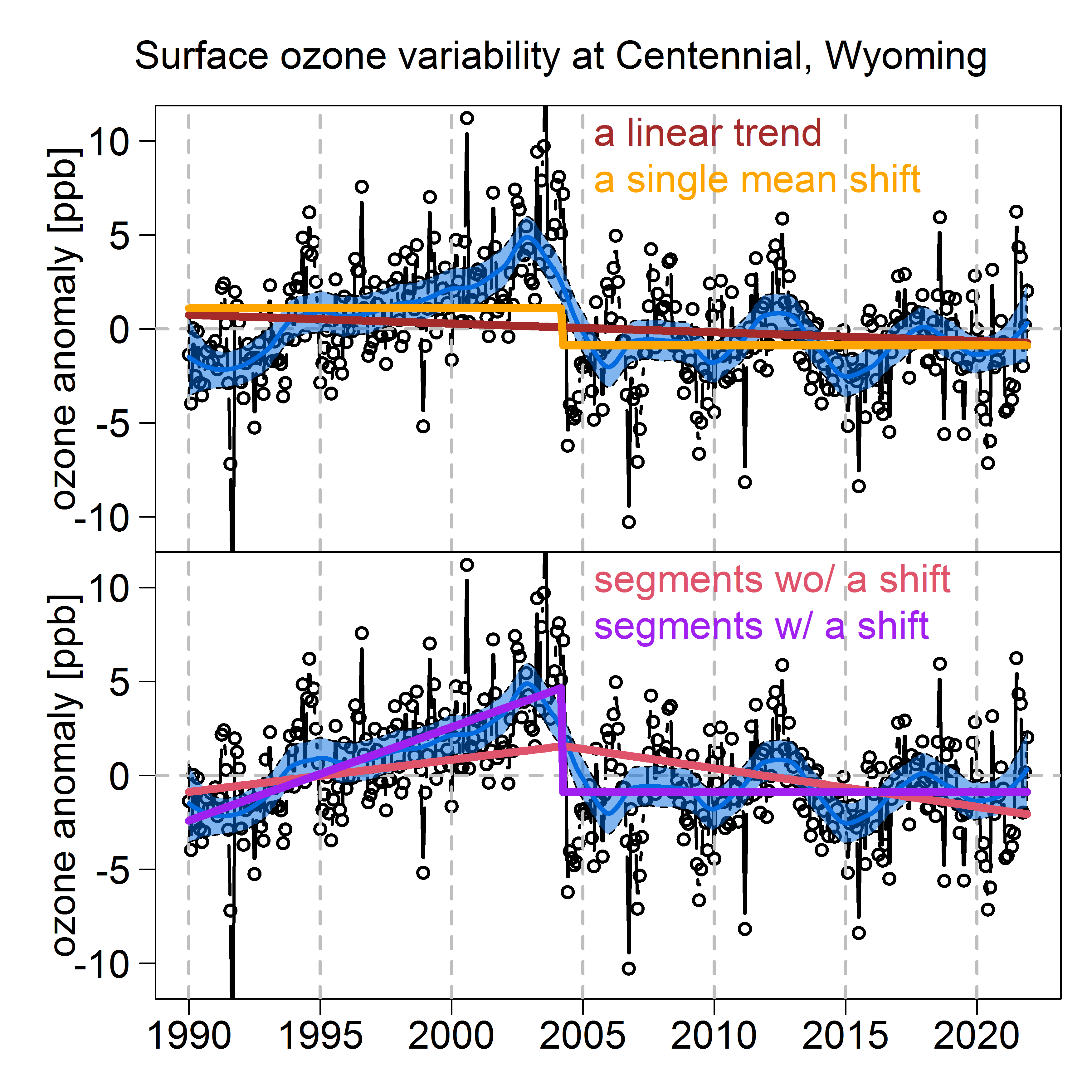}
\end{center}
 \vspace*{-5mm} 
\caption{A demonstration of change point analysis with a structural break in the time series, based on surface ozone anomalies measured at Centennial, Wyoming. The upper panel adopted a single linear trend (brown) and a single mean shift model after the structural break (orange, without considering the trends). The lower panel adopted the piecewise linear trends without (red) and with (purple) an offset in the structural break, respectively. Note that the Loess fit (blue) assumes smooth and continuous transition between data points, so it cannot handle an abrupt discontinuity properly.}
\end{figure*}

We further fit piecewise segments with a turning point (without an offset) in April 2004. This model finds that there is indeed a change point of trends, with a positive trend of 1.71 [$\pm$1.64] ppbv/decade ($p=0.04$, high certainty) over 1990-2004 and a negative trend of -2.05 [$\pm$1.27] ppbv/decade ($p \leq 0.01$, very high certainty) over 2004-2021 (dSNR = 5.3 and the 95\% confidence intervals of pre-trend and post-trend do not overlap). However, this model is still inadequate because the positive trend before the change point is obviously underestimated and the negative trend after the change point is clearly induced by the abrupt change. Under the circumstance when a change point is considered to be both a mean shift and a trend change, a piecewise trend with an offset at the change point should be adopted. This result can be summarized as a positive trend of 4.98 [$\pm$1.60] ppbv/decade ($p \leq 0.01$, very high certainty) over 1990-2004 and 0.01 [$\pm$1.15] ppbv/decade ($p=0.99$, no evidence) over 2004-2021 (dSNR = 6.2 and no overlap of the 95\% confidence intervals), with a shift of -5.58 [$\pm$0.88] ppbv ($p \leq 0.01$) at April 2004. This example demonstrates how different model assumptions can lead to diverse conclusions, thus model diagnostics and inspection are crucial for the change point analysis.

\section{Prior data quality control}
Data quality control (QC) is essential for ensuring that any analysis results are credible. Since ozone and its precursors are highly variable, automated QC methods based on purely numerical criteria might not be fully reliable. Practitioners are recommended to, 1) inspect the time series and data histograms across multiple years when checking the data consistency (e.g. to validate the trends or identify the abrupt change), and 2) compare the results from a collection of stations (e.g. geographically nearby or similar conditions), in order to avoid over-interpreting and over-generalizing the result from one specific site.

\section{Overview of studying trends from multiple time series}
Whereas this guidance note places the focus mainly on trend detection of a single time series, it is often desired or necessary to summarize the overall trends from a collection of multiple time series (e.g. over an extensive region). The complexity of merging multiple time series is dependent on if data are observed on a regular grid (this case is straightforward and will not be discussed further) or an irregular grid. In general, a direct aggregation of all time series observed at irregularly distributed locations should be avoided \citep{chang2021, diggle2010}. According to the data availability (i.e. the number and distribution of sites), and assuming that the region is relatively homogeneous (e.g. similar terrain and climate), the analysis of multiple time series observed at unevenly distributed locations can be classified according to the following scenarios.
\begin{itemize}
\item \textit{Sites are dense enough and have a good coverage across the region}: In this case it is adequate to simply aggregate the data onto a regular designated grid, then select (predefined) gridded locations to derive the trends.
\item \textit{Sites are abundant but heavily clustered within certain sub-regions}: Compared to the previous scenario, the result will be highly irregular even if the data are re-gridded. Under this circumstance the spatial heterogeneity needs to be explicitly accounted for, by using geostatistical modeling approaches \citep{cressie1990, stein1999} or machine learning techniques (e.g. random forests and neural networks, see \cite{berrocal2020} and \cite{betancourt2022}), which prevents the regional trend from being dominated by certain subregions. 
\item \textit{Sites are too few to determine the spatial pattern}: This scenario is often the case for free tropospheric measurements (profile data are sparse in space and infrequent in time). Due to the information being limited, deriving the overall trends in this scenario is challenging and subject to a larger uncertainty. Since observational sites are too few to establish a reasonable spatial model, alternative approaches are 1) to fill the spatial gaps by using validated model simulations \citep{delang2021, malashock2022}; and 2) to consider the common variability between different data sources to be representative of the trend. Signal decomposition or mixed-effects modeling techniques can be used to identify such common patterns \citep{chang2022, pedersen2019}. The greater the number of data sources, the easier the overall signal can be identified.
\end{itemize}

It should be emphasized that the above discussion is focused on how to properly combine multiple time series, but not meant to explicitly address the issue of data compatibility. For example, time series observed at nearby locations could have very different trends and uncertainties (due to localized variability and spatial heterogeneity). To address this issue, we should no longer consider this merely a time series problem, but within the scope of a longitudinal study (referring to a research design that involves repeated observations of a group of individuals over a period of time). This type of study aims to study group behavior, but individual differences are not unexpected (individuals respond varyingly, but the overall tendency can be quantified). The class of mixed-effects models in statistics is designed for the purpose of longitudinal data analysis, by including a component to quantify the group behavior (known as the fixed effect), and another component to address the individual behavior (known as the random effect). Figure A3 demonstrates the analysis result from the linear mixed effects modeling approach, by showing how to use random effects to adjust the variability from individual data sources, in addition to the overall trend. More in-depth discussions are provided by \cite{diggle2002} or introductory articles \citep{Fitzmaurice2008, harrison2018}. 

In terms of the TOAR analyses, if practitioners have doubts when combining potentially heterogeneous time series, instead of aggregating data (we cannot be sure what information will be lost after data aggregation), we recommend applying a mixed-effects modeling approach to account for all available data sources, and to allow for flexibility when considering individual data sources or local variability \citep[see examples from][]{chang2017, chang2022}.

\begin{figure*}[h!]
\begin{center}
\includegraphics[width=0.7\textwidth]{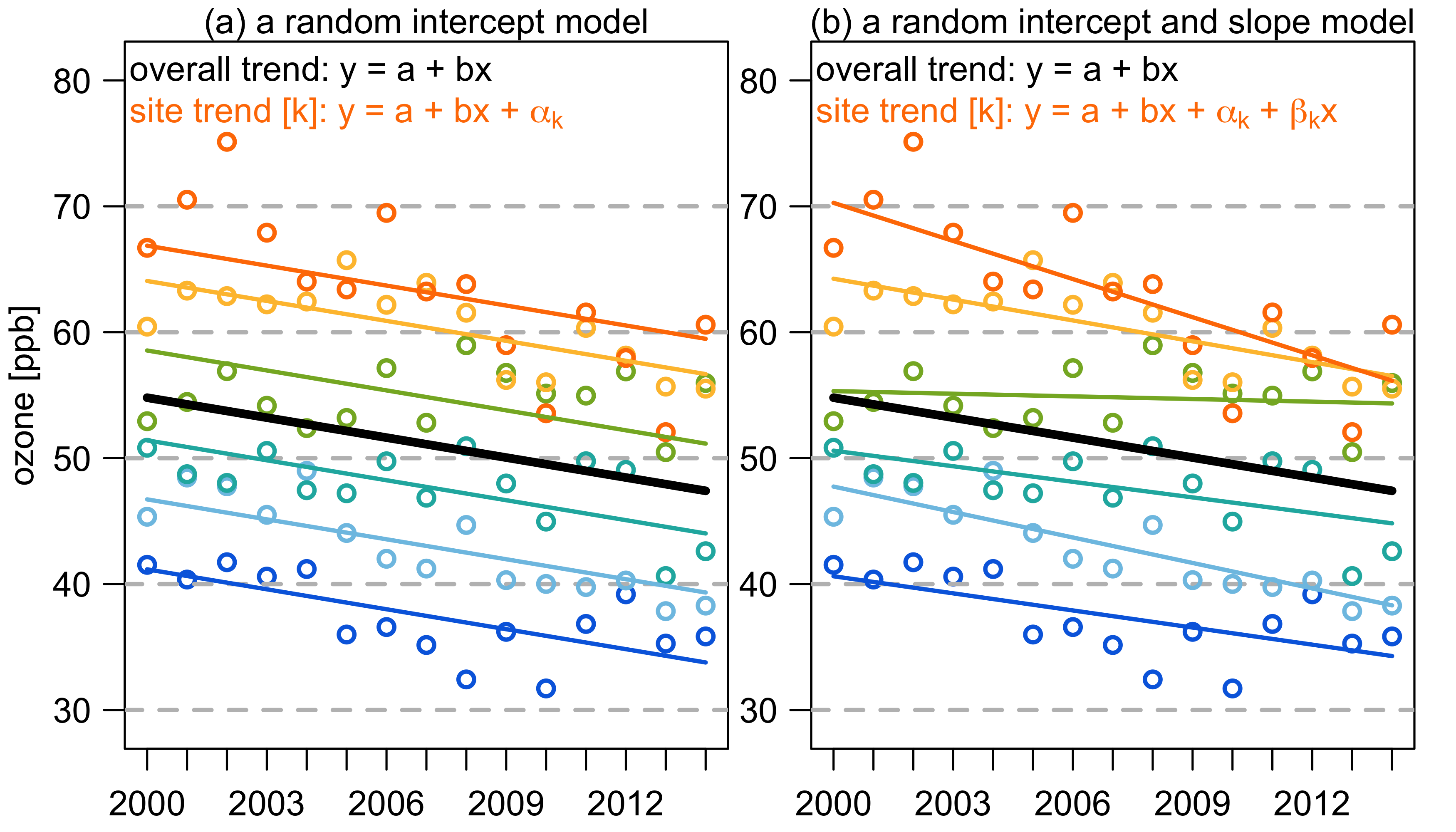}
\end{center}
\caption{A demonstration of the linear mixed-effects models: Six annual time series (averaged maximum daily 8-h averages limited to the warm season) are shown in different colors. Black line represents the overall intercept and slope (the fixed effect). Two different random effects are specified: (a) each site shares the same slope, but a site-specific intercept is adjusted to each site, and (b) a site-specific intercept/slope is adjusted to each site, so each site has its flexibility to reveal different trends. Note that the overall intercept and slope are considered to be representative, because the random intercepts and slopes from all sites are expected to have a zero mean and follow a normal distribution. Data are selected for the purpose of demonstration only (see Section 6 of \cite{chang2021} for utilizing all available sites).}
\end{figure*}

\section{Workflow process for trend analysis using quantile regression and moving block bootstrap}
This section provides a step-by-step demonstration of trend analysis using quantile regression (with Python and R code). For a data presentation, Fig A4 shows the monthly nighttime ozone time series measured at Mauna Loa, Hawaii (data and code can be downloaded at \url{https://github.com/Kai-LanChang/statistical_guidelines}). 

\begin{figure*}[h!]
\begin{center}
\includegraphics[width=0.3\textwidth]{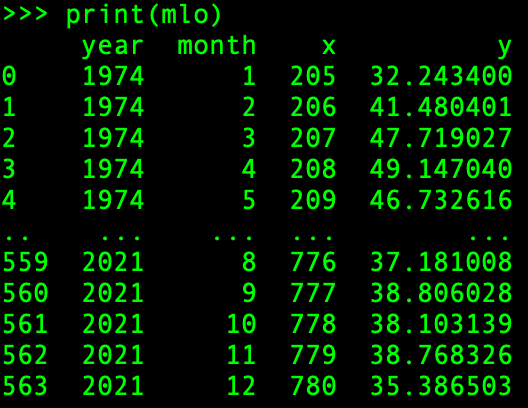}
\end{center}
 \vspace*{-7mm} 
\caption{Monthly nighttime ozone time series ($y$) measured at Mauna Loa, Hawaii.}
\end{figure*}

For the first step, we demonstrate how to deseasonalize the data. The most widely applied approach is to account for all components in a single model fit (i.e. seasonality is derived based on all available data). However, it is often desirable to use a common baseline period when we compare trends at multiple sites (or one might have greater confidence in data quality and coverage over a particular period). Therefore, in this example we demonstrate the deseasonalization based on the seasonality derived from the period 2000-2020. The seasonality is modeled as a sine-cosine combination with periodicities of 12 and 6 months.

\begin{center}
\includegraphics[width=0.7\textwidth]{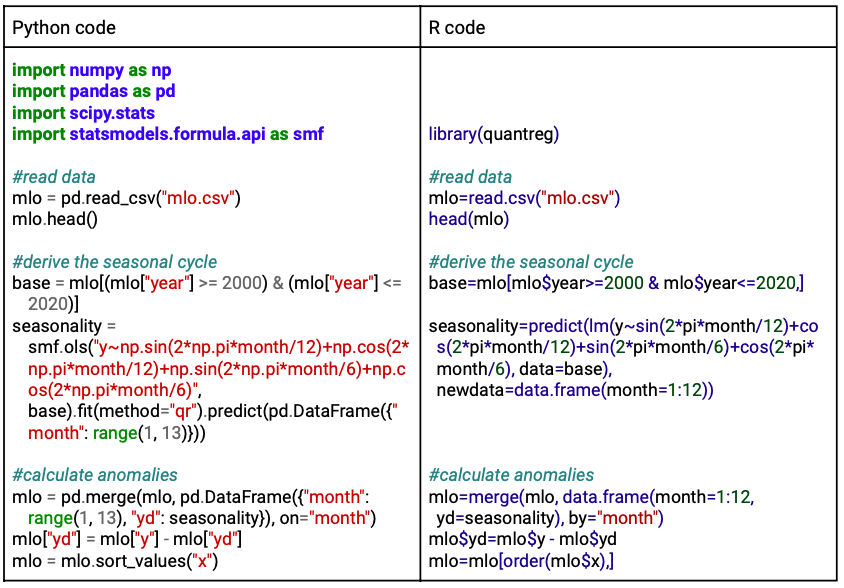}
\end{center}
 \vspace*{-7mm} 
{\footnotesize
\begin{enumerate}
  \setlength\itemsep{0em}
\item Four harmonic terms are used to capture the seasonality, these terms are subject to adjustment depending on data complexity.
\item Data needs to be sorted by the temporal index ($x$), so we can sample the data in a correct order. In this data set the monthly temporal index is referenced to January 1957 (it can be defined by users).
\end{enumerate}}
 \vspace*{6mm} 

\begin{figure*}[h!]
\begin{center}
\includegraphics[width=0.4\textwidth]{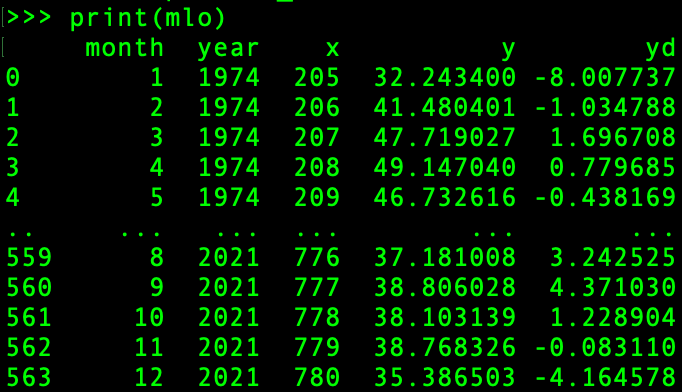}
\end{center}
 \vspace*{-7mm} 
\caption{Monthly nighttime ozone time series ($y$) and anomalies ($yd$) at Mauna Loa, Hawaii.}
\end{figure*}

The second step is to perform the quantile regression and apply a moving block bootstrap algorithm, so that the uncertainty can be properly estimated. The resulting trend value and its 1-sigma uncertainty can be found in `fit' and `fit\_se' (in units of ppbv/year). Note that typically the result is reported by ``trend [$\pm$2-sigma uncertainty]".

\begin{center}
\includegraphics[width=0.7\textwidth]{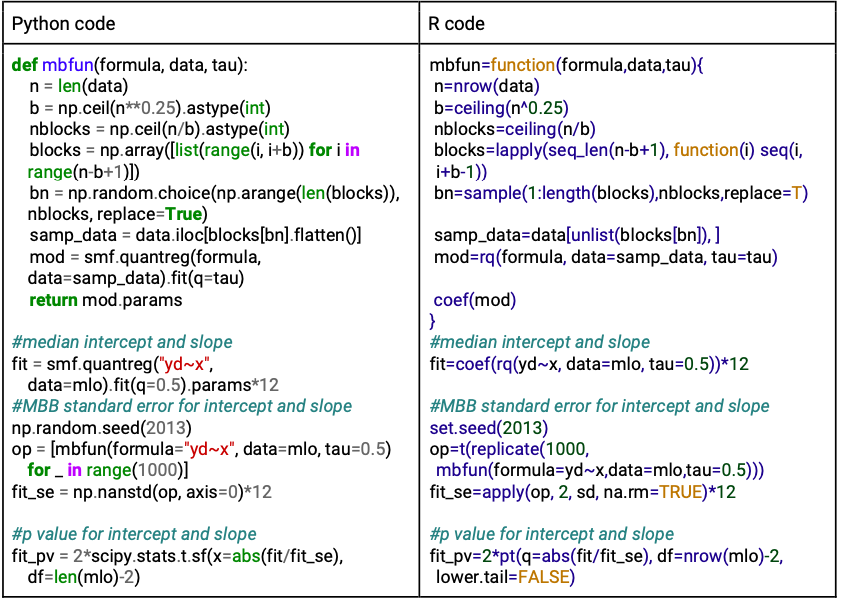}
\end{center}
 \vspace*{-7mm} 
{\footnotesize
\begin{enumerate}
  \setlength\itemsep{0em}
\item The optimal block length suggested by \cite{hall1995} is b=$n^k$, where k=1/3, 1/4 or 1/5. An alternative approach to determine the optimal block length based on autocorrelation and sample size is proposed by \cite{politis2004}.
\item Different trend techniques can be adopted by this algorithm by replacing \lstinline{quantreg()}/\lstinline{rq()} with other valid models.
\item \lstinline{q}/\lstinline{tau} is ranged between (0,1) representing the quantile to be estimated, here 0.5 (median) is used.
\item Multiple quantiles can be estimated in R at once by \lstinline{tau=seq(0.1,0.9,by=0.1)}.
\item \lstinline{random.seed()} in Python and \lstinline{set.seed()} in R create random numbers which can be reproduced (otherwise the bootstrapped results will not be the same every time the bootstrap is implemented). Note that even if the same seed number is set, random numbers are generated differently between Python and R. 
\item The sigma value is estimated based on 1000 sampled trends. 
\item The final trend and sigma values are converted from the units of ppbv/month to ppbv/year.
\item $p$-value for Student's t distribution is obtained by providing signal-to-noise ratio and degrees of freedom.
\end{enumerate}}
 \vspace*{6mm} 
 
It is worth emphasizing that this code was written for a demonstrative purpose, and that data aggregation is not a necessary step for QR. Figure A6 shows the daily and monthly ozone time series at MLO, along with the monthly 95th and 5th percentiles derived from daily values. This figure clearly demonstrates how the (potentially valuable) information is lost after monthly data aggregations. Since the extreme ozone events are of great interest to the research community, we recommend preserving those extreme variabilities, and then utilize QR for detecting heterogeneous ozone changes.

\begin{figure*}[h!]
\begin{center}
\includegraphics[width=0.7\textwidth]{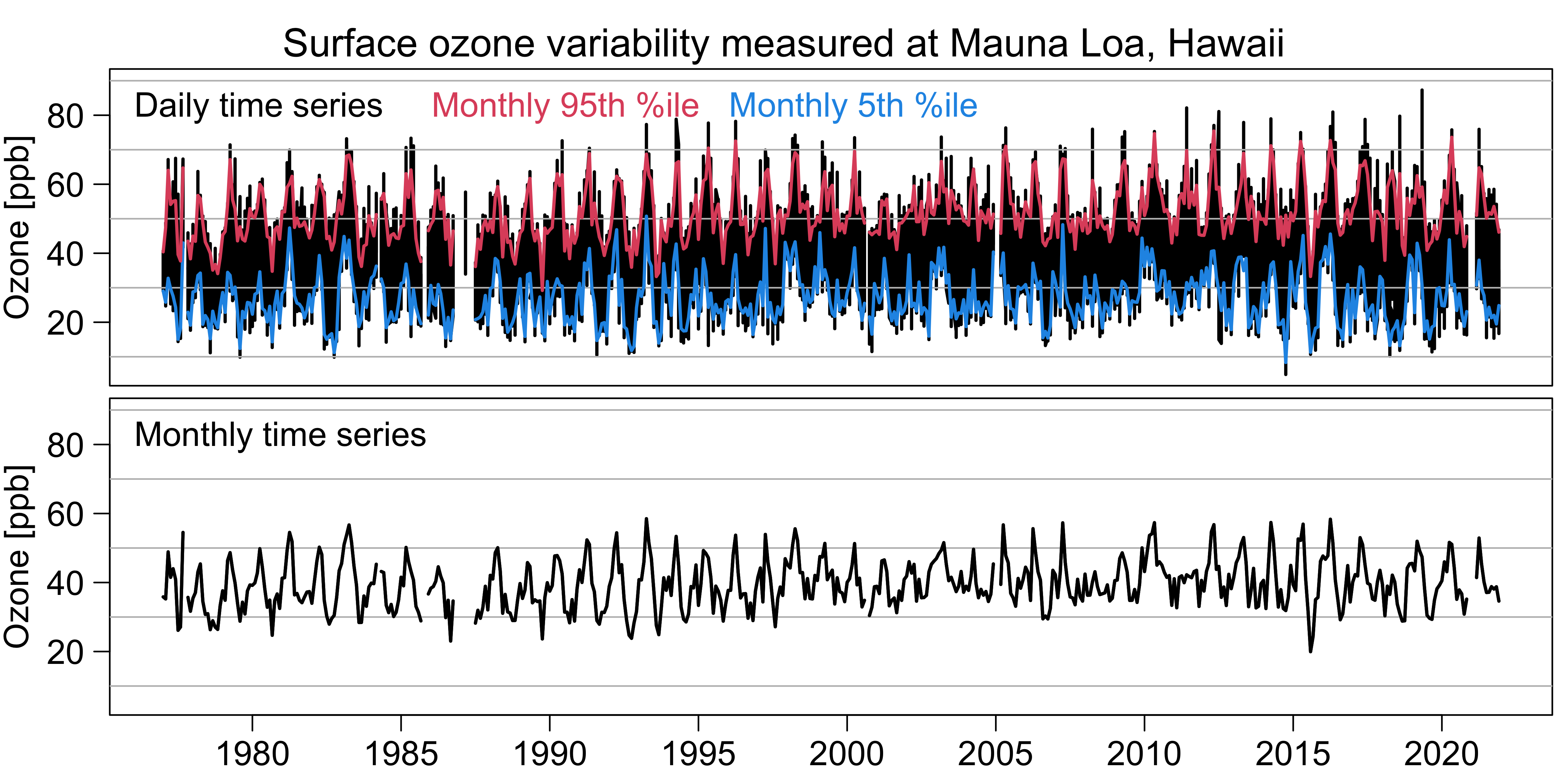}
\end{center}
 \vspace*{-5mm} 
\caption{Daily and monthly surface ozone variability at MLO, along with the monthly 95th and 5th percentiles derived from daily values.}
\end{figure*}

\section{Practical issues}
\subsection*{Different approaches to derive seasonal cycles and anomaly series }
In general, different approaches to derive the regular seasonal cycle have a minor impact on the estimation of trends \citep{chang2023, weatherhead1998}. In Section E we use the harmonic functions to approximate the seasonality, which is a typical approach for linear regression models. Another common approach for deriving seasonality is calculated by the long-term mean of each individual month (i.e. the climatology). Then the anomaly is yielded by taking the difference between the long-term mean and observed monthly value. Those two approaches might or might not yield similar seasonal cycles. For example, Figure A7 shows the seasonal cycles derived by harmonic functions and climatological values from the free tropospheric ozonesonde data measured at Boulder, Colorado and Trinidad Head (THD), California. We can see the results are nearly identical for the THD data, but some discrepancies are observed above Boulder (two modes v. a single mode). Even though the trend estimate should not be adversely affected, should the purpose be to investigate the seasonality, practitioners need to evaluate if those discrepancies are real or an artifact due to an irregular sampling scheme. Two modes of variation above Boulder are likely to be real because the result is based on a long-term record (1995-2021) with a stable sampling frequency. Multimodality is difficult to capture by lower order harmonic functions (model fits in linear regression are typically assumed to be smooth with less wiggles), but it can be adopted by using the penalized spline functions from more sophisticated generalized additive models \citep{chang2017}. Nevertheless, in terms of trend analysis both approaches should be practical.

\begin{figure*}
\begin{center}
\includegraphics[width=0.7\textwidth]{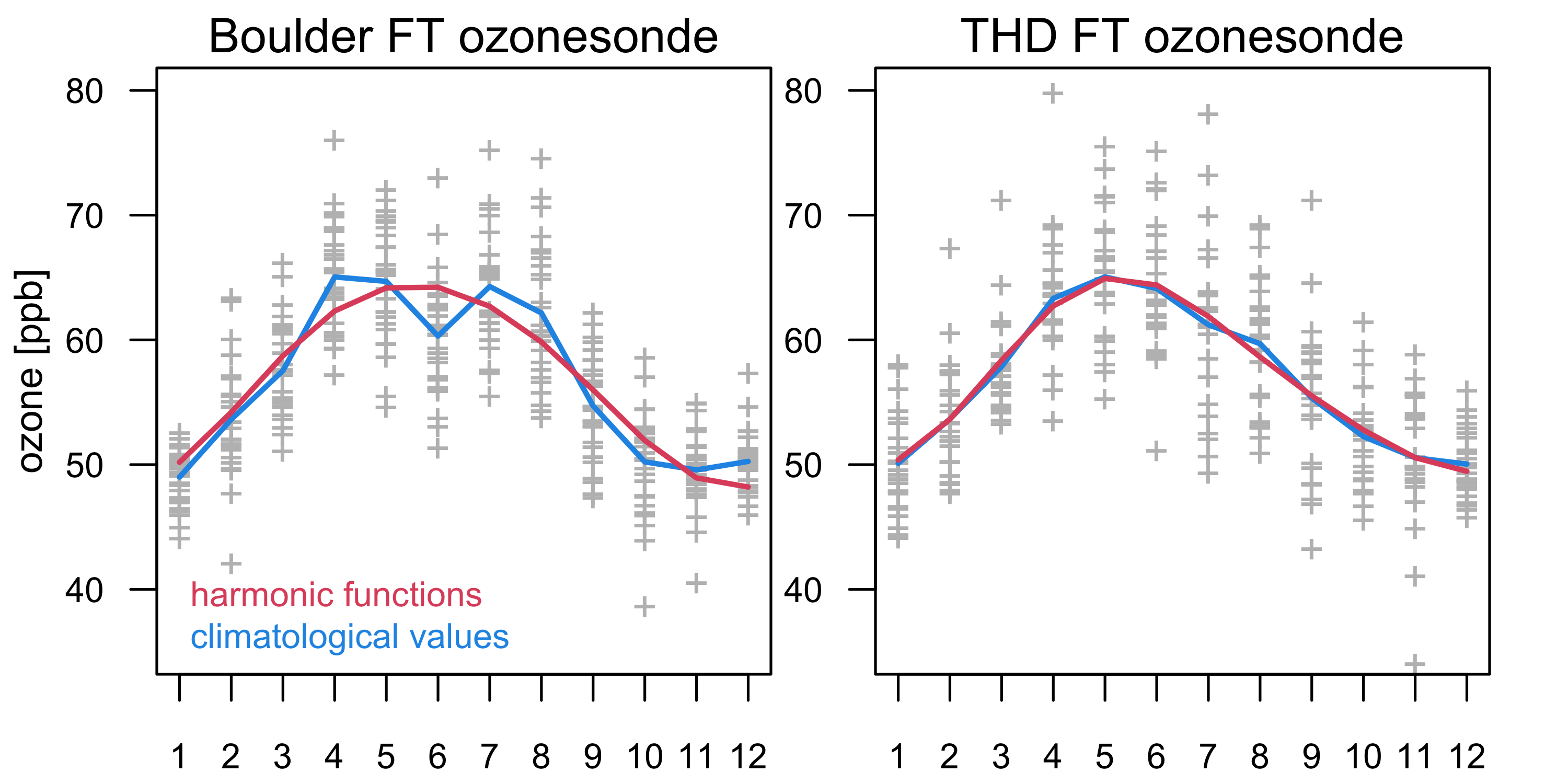}
\end{center}
 \vspace*{-5mm} 
\caption{A demonstration of different approaches to derive seasonal cycles, using harmonic functions (red) and long-term mean of each individual month (blue). Data are taken from the Boulder (Colorado, 1995-2021) and Trinidad Head (California, 1997-2021) ozonesonde records in the free troposphere (700-300 hPa).}
\end{figure*}

\subsection*{An incorporation of covariates into multiple linear quantile regressions}
Similar to conventional multiple linear regression analysis, an incorporation of covariates into QR models is possible. In this section we use several meteorological variables (dewpoint, relative humidity, wind direction, and wind speed) to fit daily surface ozone variability measured at MLO (1977-2021). The fit can be specified through the following R code:
\begin{lstlisting}
    formula=ozone~ind+dewpoint+relative_humidity+wind_direction+wind_speed+
        sin(2*pi*month/12)+cos(2*pi*month/12)+sin(2*pi*month/6)+cos(2*pi*month/6)
    mod=rq(formula, data, tau=c(0.05,0.5,0.95))
\end{lstlisting}
The fitted results are shown in Table A1. We find that meteorological variables play different roles at different percentiles: dewpoint and wind speed show a strong correlation at all three percentiles (but wind speed has stronger influence on the low percentile than high percentile), a strong correlation with relative humidity only presents for the median ozone values, and a strong correlation with wind direction only presents at the 95th percentile. Therefore, we should expect the meteorological influence to vary at different percentiles. In terms of trend estimates, the magnitudes at the 50th and 95th percentiles are similar, but the magnitudes at the 5th percentile become stronger after the meteorological variables are accounted for (mainly due to dewpoint, not shown). Confidence intervals become narrower at all three percentiles, indicating that trend uncertainties can be attributed by the meteorological influence. 

To visualize the roles of different meteorological variables, Figure A8 shows an incorporation of dewpoint and wind speed into the fit of the 95th percentile, respectively, in addition to the basic model fit with only a trend and seasonal components. Even though both dewpoint and wind speed show a strong correlation at the 95th percentile (Table A1), we can clearly see that dewpoint is more important to explain ozone variability. 

\begin{table}
\caption{Regression coefficients for meteorological variables (purple) and trends (orange, in units of ppbv/decade) at different ozone percentiles at MLO (1977-2021). Each meteorological variable is standardized so the coefficients are comparable.}
\begin{center}
\includegraphics[width=0.7\textwidth]{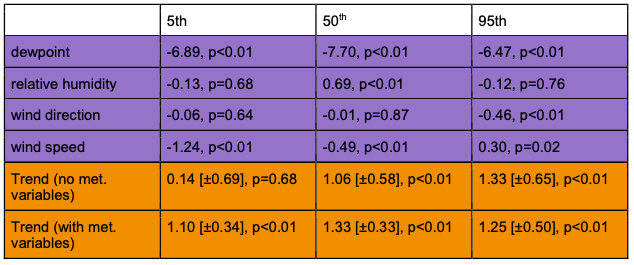}
\end{center}
\end{table}

\begin{figure*}[h!]
\begin{center}
\includegraphics[width=0.7\textwidth]{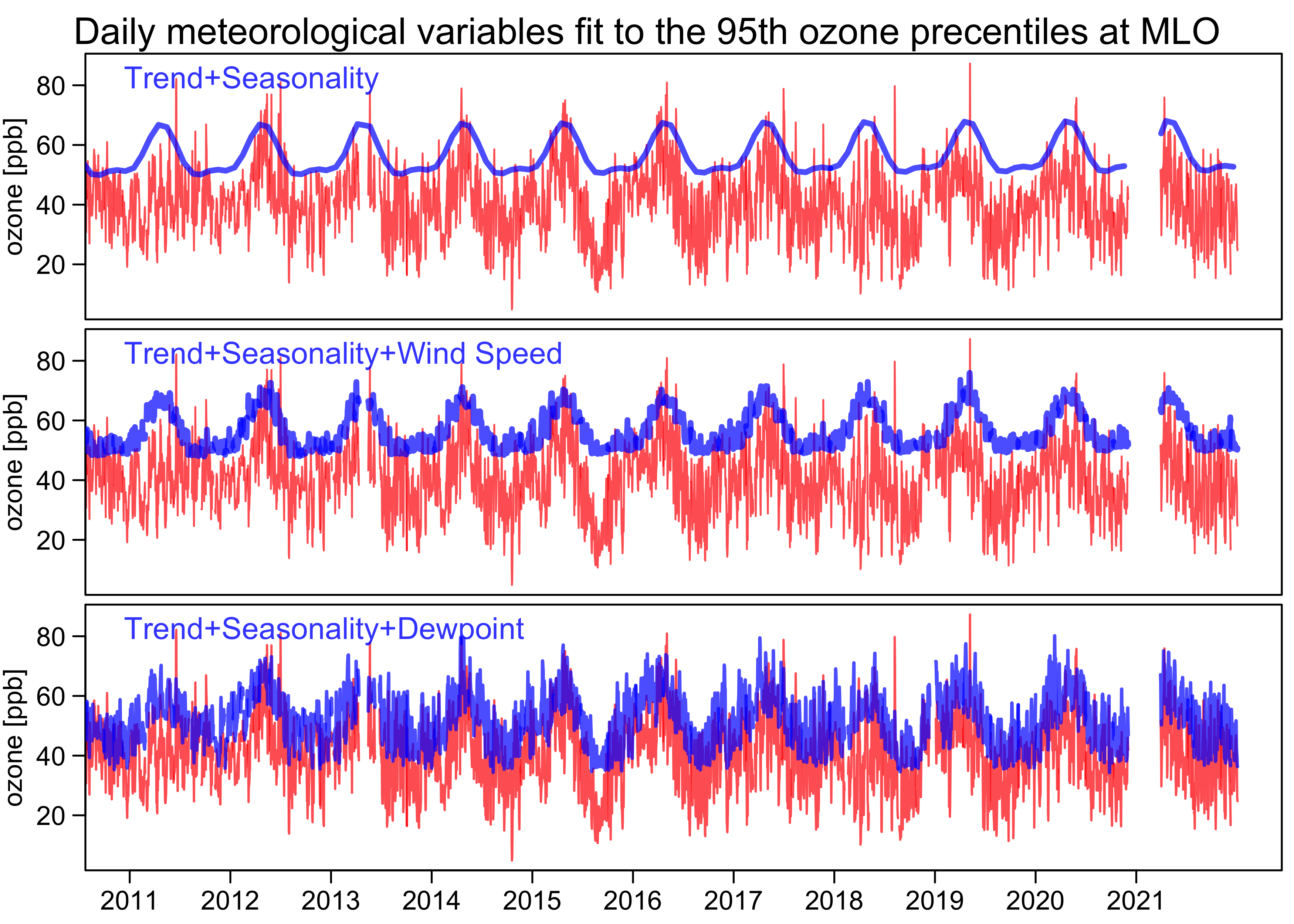}
\end{center}
 \vspace*{-5mm} 
\caption{Daily ozone time series (red) and QR fits with different model components (blue) at the 95th percentiles at MLO. Purple in the lower panels indicates observed and fitted values are overlapped. For a demonstrative purpose, only the results after 2011 are shown.}
\end{figure*}     

\subsection*{Interpolation of irregularly distributed observations onto a regular grid}
Although there are many methods for fitting a spatial interpolation model, the rationale is the same: fitting a spatial model to irregularly distributed observations for the purpose of predicting interpolated values on a regular grid. For example, Figure A9 shows the locations of monitoring stations over the southwestern USA, and we aim to conduct the trend analysis for a grid box 38-40$^\circ$N and 122-124$^\circ$W. The example code first defines the targeted grid box, and sets a loop to perform spatial interpolation for each year. The method adopts a spatial Gaussian process (\lstinline{gp}) model fitted through a generalized additive model \citep[\lstinline{gam},][]{wood2001}. Trend analysis can then be carried out on the resulting output.

\begin{figure*}
\begin{center}
\includegraphics[width=0.7\textwidth]{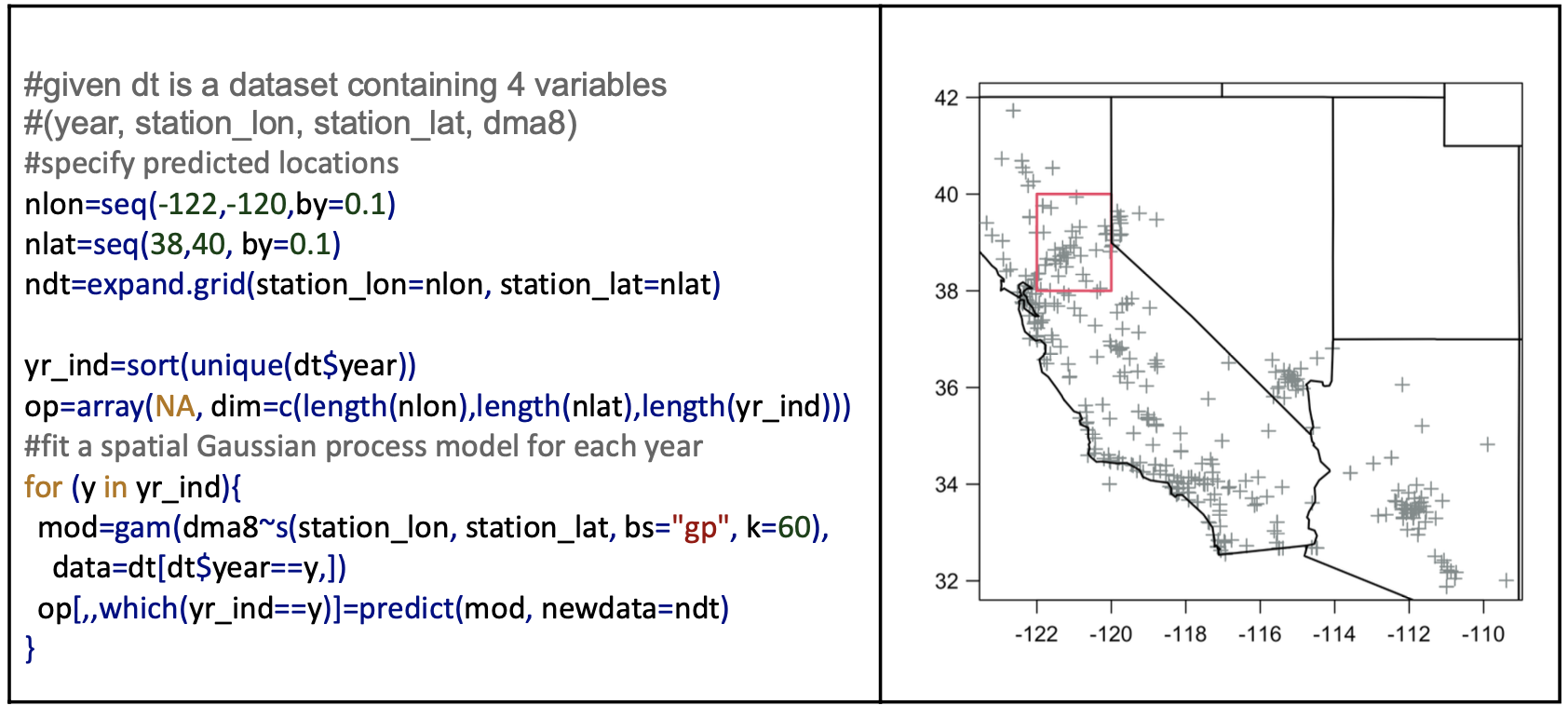}
\end{center}
 \vspace*{-5mm} 
\caption{Locations of monitoring stations (crosses) and gridded area to be predicted (red rectangle).}
\end{figure*}     

\subsection*{Impact of lagged ENSO (El Ni\~{n}o-Southern Oscillation) correlations on ozone trends}
The correlation between tropospheric ozone and ENSO is well recognized, especially across the Pacific Ocean. Its degree of correlation is varying not only with different locations, but also with different time lags. This section aims to investigate the difference between the peak and zero-lag ENSO correlations and quantify its impact on trends, by using the monthly gridded tropospheric column ozone observations measured from the OMI/MLS satellite instrument (\cite{ziemke2006, ziemke2019}, data can be found at \url{https://acd-ext.gsfc.nasa.gov/Data_services/cloud_slice/}; the ENSO index used here is taken from \url{https://psl.noaa.gov/enso/mei/};). To facilitate discussion, we use `optimized lag' to refer to the lag where the correlation peaks. The lag correlations can readily be produced by the cross-correlation function (\lstinline{ccf}) in R and Python. For a demonstration, Figure A10 shows the correlation between ENSO and ozone at different monthly lags at a grid cell 57.5$^\circ$E and 2.5$^\circ$S. For this particular grid cell the optimized lag is located at -4, indicating that a strong ENSO signal might be highly correlated with ozone measured 4 months later.

We proceed by extracting zero-lag correlation and the peak correlation between $\pm$6 monthly lags at each cell (a longer correlation range is not considered), and the resulting maps are shown in Figure A11. For the regions with the strongest ENSO correlations (i.e. tropical Pacific Ocean and Southeastern Asia), we find that the correlation patterns are highly similar between two scenarios. Nevertheless, the grid cells with a very weak zero-lag correlation in the Northern Hemisphere and Tropics generally show a stronger positive correlation after the optimized lag is found, with some exceptions in the eastern North Atlantic Ocean where negative correlation is present. The median trend estimates are shown in Figure A12, three scenarios are considered here: 1) no ENSO is taken into account, 2) zero-lag ENSO correlation is incorporated, and 3) the optimized lag ENSO correlation is incorporated. Even if some minor discrepancies can be identified, highly consistent patterns can be observed between these three scenarios. 
     
To explicitly quantify the ENSO impact, we define the relative change of a value A versus a value B as $(A - B)/B$. We calculate the relative change in trends and uncertainties for zero-lag v no ENSO correlations (case 1), and the peak v zero-lag ENSO correlations (case 2), respectively, and the result is shown in Figure A13.  The only systematic bias is found from the relative change in trend uncertainties in case 1. After the ENSO correlation is accounted for, the uncertainties become smaller over the tropical Pacific Ocean (where the strongest ENSO correlation is located), indicating that certain trend uncertainty can be attributed by the ENSO signal. For the relative change in trends, 95.8\% cells in case 1 and 97.0\% cells in case 2 fall within $\pm$0.5\% bias, and 98.8\% cells in case 1 and 98.4\% cells in case 2 fall within $\pm$1\% bias. The relative change in uncertainties are all limited within 0.7\% bias for both cases. In summary, in this case study we may conclude that an incorporation of ENSO in trend analysis could reduce trend uncertainty and avoid potential systematic bias at certain regions, but its impact on trend estimate is expected to be weak. Even though the peak correlation between ENSO and ozone might be temporally lagged, it does not necessarily have an impact on trends and uncertainties.  The complete R code to produce all figures in this section is provided at \url{https://github.com/Kai-LanChang/statistical_guidelines}.    

\begin{figure*}
\begin{center}
\includegraphics[width=0.3\textwidth]{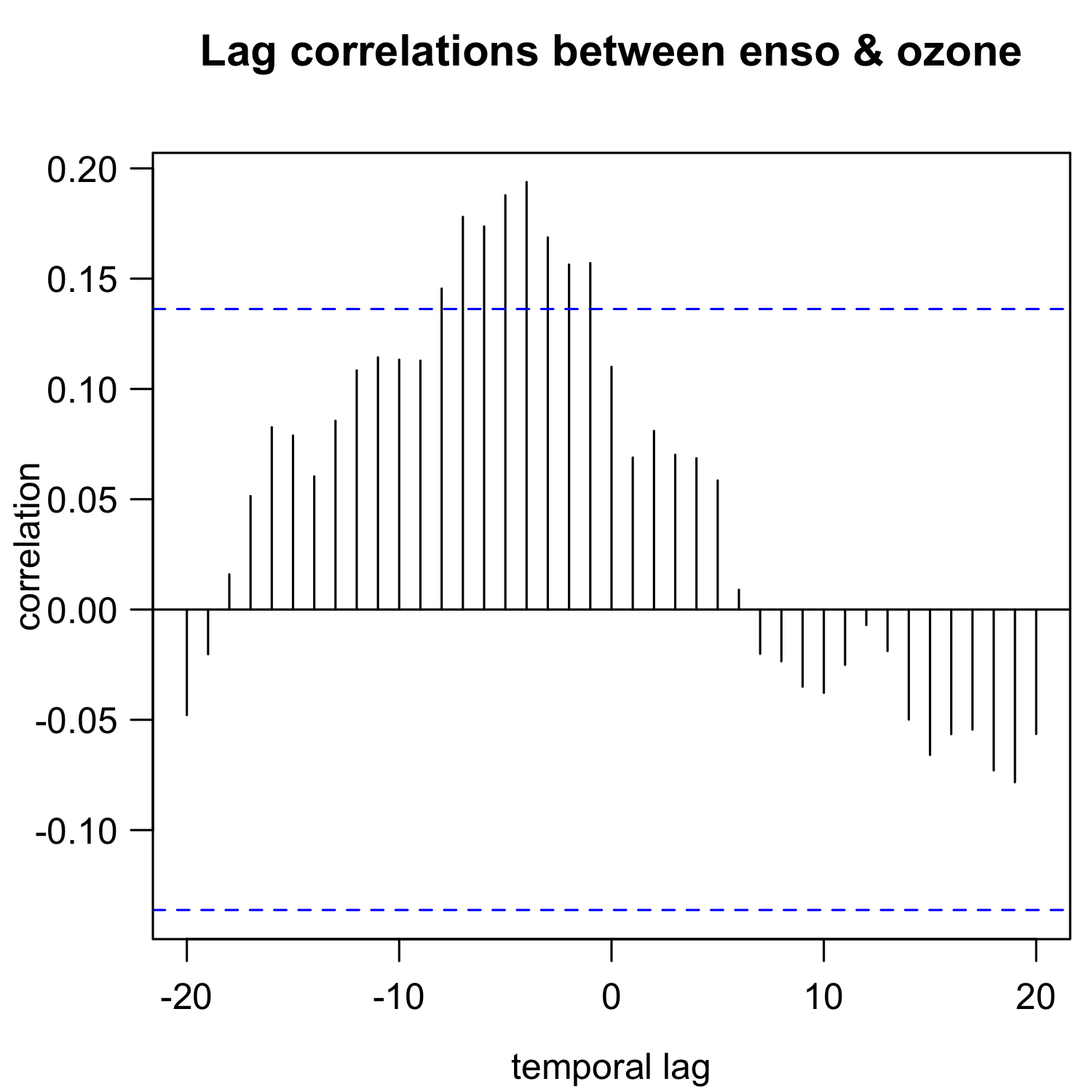}
\end{center}
 \vspace*{-5mm} 
\caption{Correlation between ENSO and ozone at different monthly lags (OMI/MLS data, 57.5$^\circ$E and 2.5$^\circ$S). The peak (positive) correlation is found at -4 lag.}
\end{figure*}     

\begin{figure*}
\begin{center}
\includegraphics[width=0.7\textwidth]{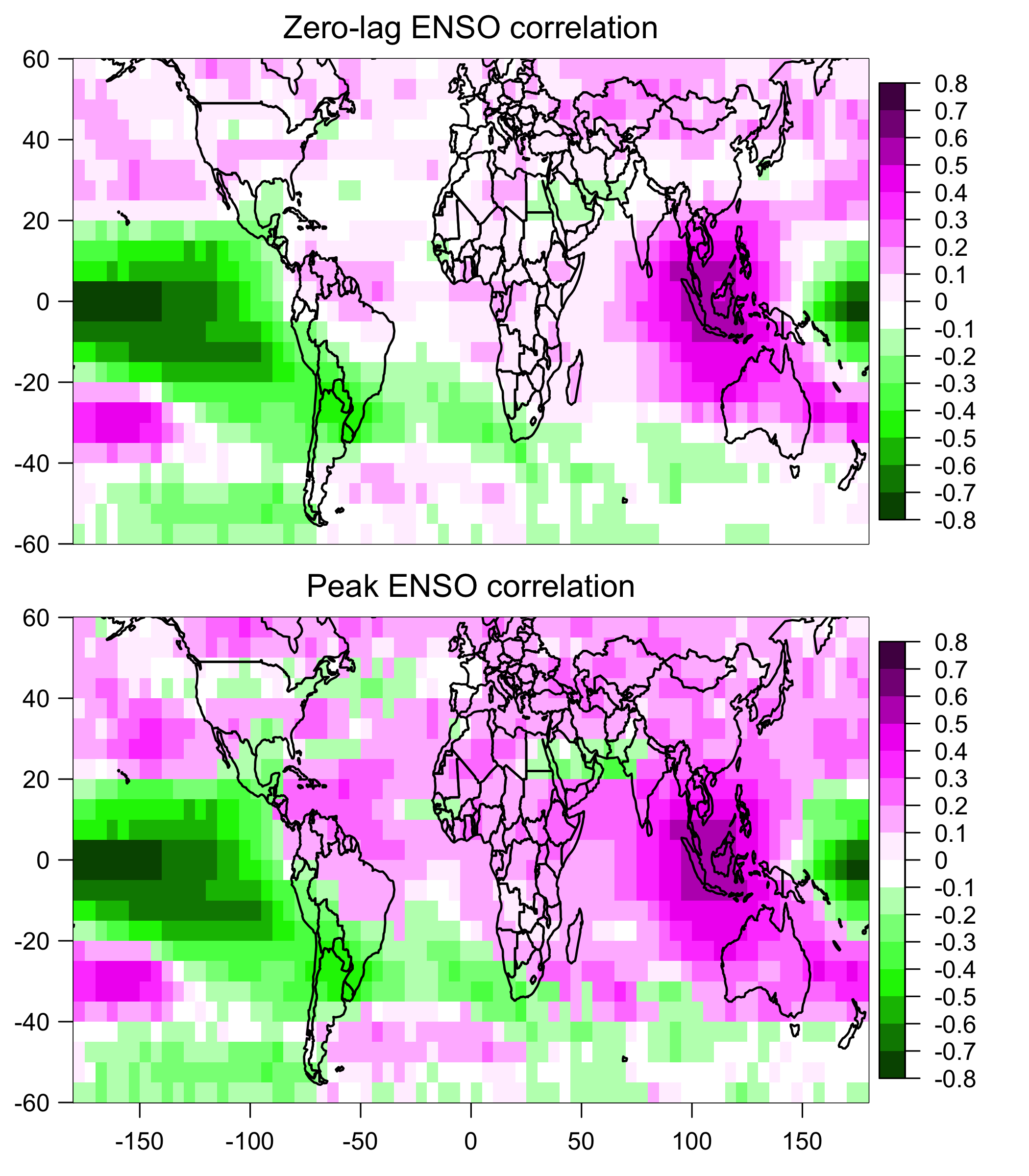}
\end{center}
 \vspace*{-5mm} 
\caption{Correlation coefficients between ENSO and ozone from the OMI/MLS dataset (2004-2021): the upper panel shows zero-lag ENSO correlation, and the lower panel shows the peak ENSO correlation between $\pm$6 monthly lags.}
\end{figure*}     

\begin{figure*}
\begin{center}
\includegraphics[width=0.7\textwidth]{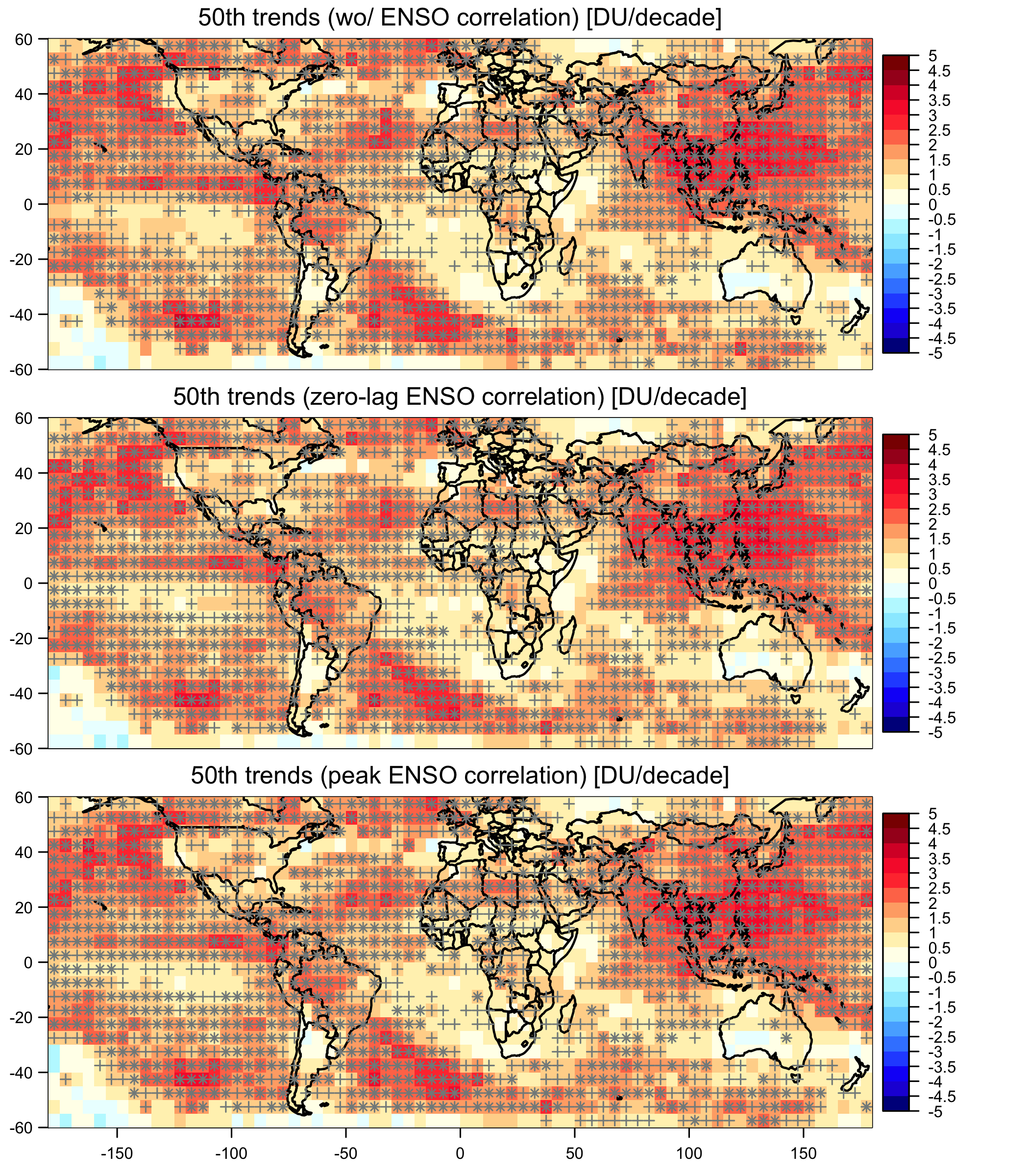}
\end{center}
 \vspace*{-5mm} 
\caption{Median trend estimates [DU/decade] from the OMI/MLS dataset (2004-2021): the upper panel shows the estimation without incorporating ENSO correlation, the middle panel is based on the zero-lag ENSO correlation, and the lower panel is based on the peak ENSO correlation between $\pm$6 monthly lags. Gray crosses indicate grid cells with a $p$-value between 0.01 and 0.05, and stars indicate grid cells with a $p$-value less than or equal to 0.01.}
\end{figure*}     

\begin{figure*}
\begin{center}
\includegraphics[width=0.7\textwidth]{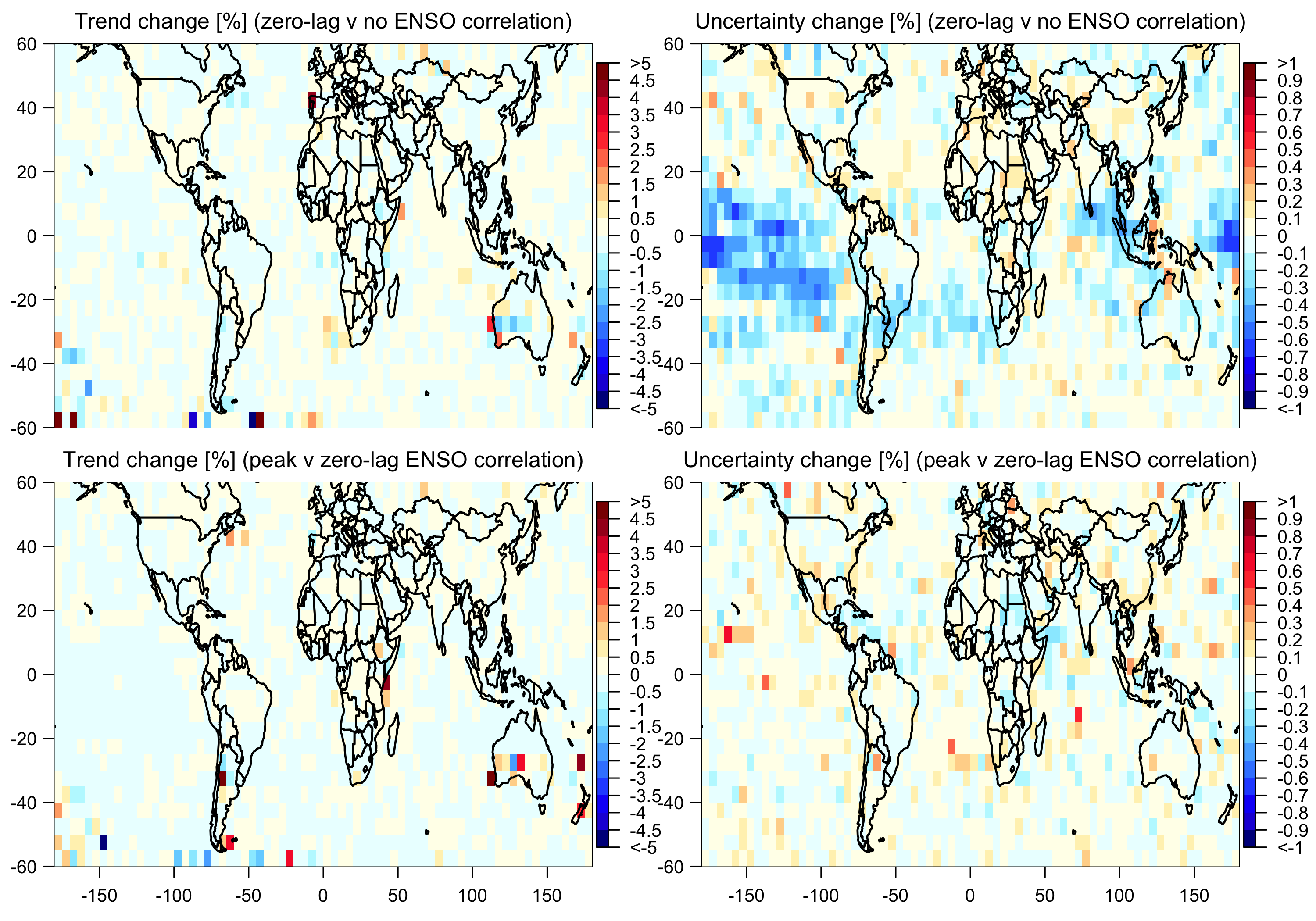}
\end{center}
 \vspace*{-5mm} 
\caption{Relative change [\%] in median trends (left) and uncertainties (right) for zero-lag v no ENSO correlations (top), and the peak v zero-lag ENSO correlations (bottom), respectively.}
\end{figure*}

\bibliographystyle{apa3}
{\small \bibliography{STAT_recommendations_TOAR_analyses}}
\end{document}